\documentclass[prx,superscriptaddress,floatfix,showpacs,notitlepage,twocolumn]{revtex4-2}
\usepackage[latin9]{inputenc}
\usepackage{color}
\usepackage{textcomp}
\usepackage{amsmath}
\usepackage{amssymb}
\usepackage{mathtools}
\usepackage{graphicx}
\usepackage{esint}
\usepackage{framed} 
\usepackage[dvipsnames]{xcolor}
\usepackage{hyperref}
\usepackage{enumerate}
\usepackage{enumitem}
\usepackage{tikz-network}
\usepackage{moreenum}
\hypersetup{
  linkcolor = black,
  citecolor  = blue,
  urlcolor = black,
  colorlinks = true,
}
\makeatletter

\usepackage{amsfonts}
\usepackage{bbold}
\usepackage{natbib}
\usepackage{svg}
\usepackage{float}
\usepackage[export]{adjustbox}
\hypersetup{    colorlinks,    linkcolor={blue!50!black},    citecolor={blue!50!black},    urlcolor={blue!80!black}}
\usepackage[all]{hypcap}
\usepackage{bbold}
\graphicspath{{img/}}

\newcommand*{\bra}[1]{\ensuremath{\langle #1 \vert}}
\newcommand*{\ket}[1]{\ensuremath{\vert #1 \rangle}}

\newcommand*{\tr}[1]{\mathrm{tr}\left(#1\right)}

\newcommand{\mc}[1]{\mathcal{#1}}

\newcommand{\lr}[1]{\left( #1 \right)}

\renewcommand{\vec}[1]{{\boldsymbol{#1}}}

\begin{document}

\title{{Protocols for Rydberg entangling gates featuring  robustness against quasi-static errors}}

 \author{Charles Fromonteil}
 \affiliation{Institute for Theoretical Physics, University of Innsbruck, 6020 Innsbruck, Austria}
 \affiliation{Institute for Quantum Optics and Quantum Information of the Austrian Academy of Sciences, 6020 Innsbruck, Austria}
\author{Dolev Bluvstein}
\affiliation{Department of Physics, Harvard University, Cambridge, MA02138, USA}
 \author{Hannes Pichler}
 \affiliation{Institute for Theoretical Physics, University of Innsbruck, 6020 Innsbruck, Austria}
 \affiliation{Institute for Quantum Optics and Quantum Information of the Austrian Academy of Sciences, 6020 Innsbruck, Austria}

\begin{abstract}

We introduce a novel family of protocols for entangling gates for neutral atom qubits based on the Rydberg blockade mechanism. These protocols realize controlled-phase gates through a series of global laser pulses that are on resonance with the Rydberg excitation frequency. We analyze these protocols with respect to their robustness against calibration errors of the Rabi frequency or shot-to-shot laser intensity fluctuations, and show that they display robustness in various fidelity measures. In addition, we discuss adaptations of these protocols in order to make them robust to atomic-motion-induced Doppler shifts as well.
\end{abstract}

\maketitle
\section{Introduction}
Arrays of neutral atoms trapped by optical tweezers provide a powerful platform for implementing quantum information processing protocols \cite{browaeysManybodyPhysicsIndividually2020,weimerRydbergQuantumSimulator2010a,bernienProbingManybodyDynamics2017a,labuhnTunableTwodimensionalArrays2016,
jakschFastQuantumGates2000b,lukinDipoleBlockadeQuantum2001,urbanObservationrydberg,wilkEntanglementTwoIndividual2010_2,levineParallelImplementationHighFidelity2019,kaufmanQuantumScienceOptical2021,grahamMultiqubitEntanglementAlgorithms2022,maUniversalGateOperations2022,madjarovHighfidelityEntanglementDetection2020a,bluvsteinQuantumProcessorBased2022_2,omranGenerationManipulationSchrodinger2019a,semeghiniProbingTopologicalSpin2021_2,youngHalfminutescaleAtomicCoherence2020,ebadiQuantumOptimizationMaximum2022a,byunFindingMaximumIndependent2022,bluvsteinQuantumProcessorBased2022_2,singhDualElementTwoDimensionalAtom2022,steinertSpatiallyProgrammableSpin2022_3}. A promising approach is to encode quantum information in long-lived hyperfine states (or other long-lived electronic states), with each atom representing a single qubit. Single qubits can be initialized with high fidelity via optical pumping, read out via tweezer-resolved imaging techniques, and coherently manipulated with high fidelity via optical control. Moreover, high-fidelity multi-qubit gates enabled by state-selective coherent excitation to strongly interacting Rydberg states have been realized \cite{jakschFastQuantumGates2000b,lukinDipoleBlockadeQuantum2001,urbanObservationrydberg,wilkEntanglementTwoIndividual2010_2,levineParallelImplementationHighFidelity2019,kaufmanQuantumScienceOptical2021,grahamMultiqubitEntanglementAlgorithms2022,maUniversalGateOperations2022,madjarovHighfidelityEntanglementDetection2020a,bluvsteinQuantumProcessorBased2022_2}. In combination with coherent rearrangement techniques, this results in a versatile, scalable quantum processing architecture \cite{bluvsteinQuantumProcessorBased2022_2}.%

In this work, we introduce novel protocols for realizing multi-qubit entangling gates in this platform, with several appealing features: similar to recently introduced protocols \cite{levineParallelImplementationHighFidelity2019,janduraTimeOptimalTwoThreeQubit2022,paganoErrorbudgetingControlledphaseGate2022_2}, our new protocols consist of a sequence of global laser pulses, avoiding the requirement for local addressing with Rydberg lasers. In contrast to previous proposals, all these pulses are resonant with the Rydberg transition and require pulse areas that can be simply calibrated. In addition, the protocols introduced exhibit a native robustness against calibration errors of the Rabi frequency and low-frequency or shot-to-shot fluctuations of the laser intensity. Moreover, we also discuss methods to suppress errors due to Doppler shifts from thermal motion of the atoms. Although in recent experiments the primary limitations can often be incoherent errors such as Rydberg decay and laser scattering \cite{bluvsteinQuantumProcessorBased2022_2}, with improvements in technology and laser power these quasi-static errors will become more relevant, and ultimately be important to ensure robust quantum circuit operation.

\section{Model}\label{Sec:Model}
We consider a pair of atoms, where two non-interacting stable internal states of each atom represent the two qubit states $\ket{0}$ and $\ket{1}$. We are interested in entangling gates between these qubits that are mediated via state-selective coherent excitation from the state $\ket{1}$ to an interacting Rydberg state $\ket{r}$. The Hamiltonian governing the dynamics of a pair of atoms driven by a laser that induces such a laser an excitation process is given by \cite{levineParallelImplementationHighFidelity2019}
\begin{align} \label{eq:Hamiltonian}
    H=&\sum_{i=1,2}\frac{\Omega}{2}\lr{e^{i\varphi}\ket{1}_i\bra{r}_i+\textrm{h.c.}}+V \ket{r,r}\bra{r,r}.
\end{align}
Here, $\Omega$ and $\varphi$ are the (real) Rabi frequency and the phase of the laser, respectively. We stress that we consider the situation of a homogeneous, i.e., global laser field, thus eliminating the requirement for local control. Moreover, the laser frequency is resonant with the transition between the state $\ket{1}$ and the Rydberg state $\ket{r}$. In the following we are interested in the situation where the atoms are placed at distances such that the interaction energy in the state where both atoms are in the Rydberg state, $V$, is much larger than the Rabi frequency ($V\gg |\Omega|)$. This results in a dynamical constraint that suppresses simultaneous excitation of both atoms to the Rydberg state \cite{jakschFastQuantumGates2000b,lukinDipoleBlockadeQuantum2001,saffmannQuantumInformationRydberg2010_2}. To simplify the discussion in the following we enforce this Rydberg blockade constraint exactly, which is formally equivalent to considering the limit $V\rightarrow \infty$. 

\begin{figure}[t]
    \centering
    \includegraphics[width=\linewidth]{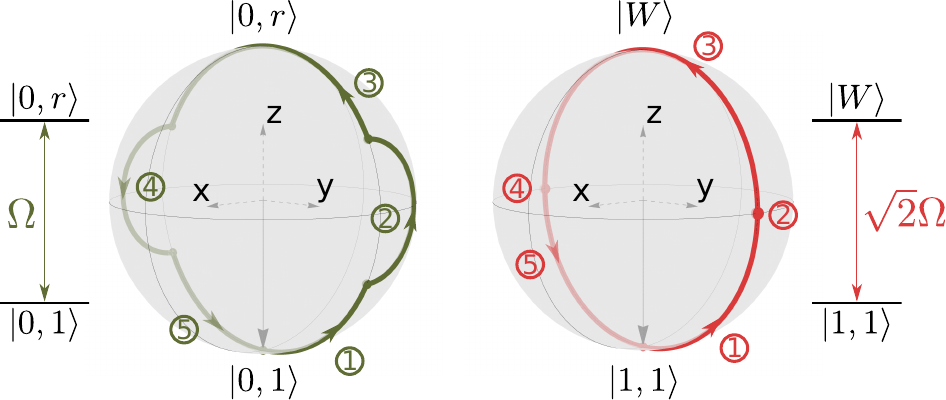}
    \caption{Bloch sphere trajectories followed by the effective two-level system $\ket{0,1}\leftrightarrow \ket{0,r}$ and equivalently $\ket{1,0}\leftrightarrow \ket{r,0}$ (left) and of the effective two-level system $\ket{1,1}\leftrightarrow \ket{W}$ (right) during the sequence \eqref{eq:variant1}. The states $\ket{0,1}$ and $\ket{1,0}$ are respectively coupled to $\ket{0,r}$ and $\ket{r,0}$ with a coupling strength $\Omega$, while $\ket{1,1}$ is coupled to $\ket{W}=\frac{\ket{1,r}+\ket{r,1}}{\sqrt{2}}$ with a coupling strength $\sqrt{2}\Omega$.
    Both trajectories are closed, with an enclosed area of $2\pi$, such that the three computational basis states $\ket{1,0}$, $\ket{0,1}$ and $\ket{1,1}$ pick up a minus sign. Since the state $\ket{0,0}$ does not evolve, this realizes a CZ gate.}
    \label{fig:traj}
\end{figure}

Owing to the global drive and the blockade constraint, the dynamics of the four computational basis states decomposes into simple blocks.
The state $\ket{0,0}$ does not couple to the light field and is trivially invariant. The state $\ket{0,1}$ is resonantly coupled to the state $\ket{0,r}$, with a coupling matrix element $\Omega e^{i\varphi}/2$, forming an effective two level system. The dynamics of the state $\ket{1,0}$ exactly mirrors this, as $\ket{1,0}$ resonantly couples to the state $\ket{r,0}$, with the same coupling strength $\Omega e^{i\varphi}/2$. Finally, the state $\ket{1,1}$ is resonantly coupled to the state $\ket{W}=\tfrac{1}{\sqrt{2}}(\ket{1,r}+\ket{r,1})$, forming again a closed, effective two-level system. However, in this last case the coupling matrix element is larger and given by $\sqrt{2}\Omega e^{i\varphi}/2$. The entire dynamics can thus be understood by considering two inequivalent effective two level systems, which can be conveniently represented by two Bloch spheres where the south pole is identified with a computational basis state and the north pole with the corresponding coupled state containing a Rydberg excitation  \cite{levineParallelImplementationHighFidelity2019}. These two Bloch spheres are depicted in Fig.~\ref{fig:traj}, where the trajectory of the Bloch vector on the left sphere depicts the dynamics of the states $\ket{0,1}$ and equivalently $\ket{1,0}$, whereas the right sphere shows the dynamics of the state $\ket{1,1}$ for a given pulse sequence specified below. The  Hamiltonian \eqref{eq:Hamiltonian} induces rotations on both Bloch spheres. While the rotation axis (determined by $\varphi$) is the same for both spheres, the rotation frequency differs by a factor $\sqrt{2}$. For future reference we define the coordinate system such that $\varphi=0$ corresponds to a rotation around the $x$ axis, while $\phi=\pm\pi/2$ corresponds to a rotation around the $\pm y$ axis (see Fig.~\ref{fig:traj}). Note that the state $\ket{0,0}$ does not evolve at all under $\eqref{eq:Hamiltonian}$.

\section{Controlled phase gate Protocol}\label{Sec:gate}

We proceed by introducing two closely related pulse sequences that result in a controlled-$Z$  gate, denoted by CZ between the two qubits. Its action on the qubit states is given by  $\textrm{CZ}\ket{z_1,z_2}= (-1)^{z_1z_2-z_1-z_2}\ket{z_1,z_2}$ with ($z_i\in \{0,1\}$). We note that this definition of the CZ gate differs from the standard one by a trivial rotation of both qubits. To realize this gate, we use a sequence of resonant laser pulses (with fixed Rabi frequency $\Omega$), each specified by the pulse duration $t$ and the phase of the laser $\varphi$, or equivalently, the rotation axis. We denote the unitaries generated by Hamiltonian \eqref{eq:Hamiltonian} from a pulse of duration $t=\alpha/\Omega$ with laser phase $\varphi=0$ by $U_{x}(\alpha)$. Analogously we denote the operations corresponding to pulses with  $\varphi=\pi$, $\varphi=\pi/2$ and $\varphi=-\pi/2$, by  $U_{-x}(\alpha)$, $U_{y}(\alpha)$ and $U_{-y}(\alpha)$ respectively. We note that we consider square pulses here for simplicity, however, due to the resonant nature of the pulses only the pulse area matters for all aspects discussed in this section (experimentally, intensity-dependent light shifts of the transition need to be appropriately calibrated. Both our primary protocols are constructed by a sequence of 5 such pulses, each of them with a pulse area $\Omega t\in\{\pi,\frac{\pi}{2},\tfrac{\pi}{\sqrt{2}},\tfrac{\pi}{2\sqrt{2}}\}$: the pulse sequences are given by
\begin{align}\label{eq:variant1}
\textrm{CZ}=U_x\lr{\tfrac{\pi}{2\sqrt{2}}}U_y(\pi)U_x\lr{\tfrac{\pi}{\sqrt{2}}}U_y(\pi)U_x\lr{\tfrac{\pi}{2\sqrt{2}}},
\end{align}
and
\begin{align}\label{eq:variant2}
\textrm{CZ}=U_x\lr{\tfrac{\pi}{2}}U_y\lr{\tfrac{\pi}{\sqrt{2}}}U_x(\pi)U_y\lr{\tfrac{\pi}{\sqrt{2}}}U_x\lr{\tfrac{\pi}{2}}.
\end{align}
Note that these pulse sequences are reminiscent of the refocusing (spin echo) techniques used in NMR \cite{VandersypenNMRtechniques,GullionCarrPurcell}.
While it is a matter of simple algebra to confirm that they both realize the target controlled-$Z$ gate, we find it instructive to consider a geometric argument obtained by analyzing the path traced out by the computational basis states on the Bloch spheres introduced above. We consider here the case of the sequence \eqref{eq:variant1}, but analogous arguments apply to \eqref{eq:variant2}. For this analysis it is convenient to formally split the third pulse in the sequence into two identical pulses, each with half the pulse area, i.e.,  $U_x\lr{\pi/\sqrt{2}}=U_x\lr{\tfrac{\pi}{2\sqrt{2}}}U_x\lr{\tfrac{\pi}{2\sqrt{2}}}$. With this, the pulse sequence \eqref{eq:variant1} can be interpreted as a sequence consisting of 6 pulses. Note that these 6 pulses are obtained by applying the same 3-pulse sequence twice: we can write the evolution operator generated by the 6 pulses as a repeated application of the evolution operator describing the first three pulses, $S$, i.e., ${\rm{CZ}}=S^2$ with $S=U_x\lr{\tfrac{\pi}{2\sqrt{2}}}U_y(\pi)U_x\lr{\tfrac{\pi}{2/\sqrt{2}}}$. The 3-pulse sequence described by $S$ has a simple effect on the four computational basis states of the two atoms: While the state $\ket{0,0}$ is trivially invariant, $S\ket{0,0}=\ket{0,0}$, all other computational basis states are simultaneously mapped from the south pole of the corresponding Bloch sphere to the north pole, i.e., $S\ket{0,1}=i\ket{0,r}$, $S\ket{1,0}=i\ket{r,0}$ and $S\ket{1,1}=i\ket{W}$ (see Fig.~\ref{fig:traj}).
This can be seen by analyzing the trajectories of each computational basis state in Fig.~\ref{fig:traj}: On the left, we show the trajectories of an atom pair initially on the state $\ket{0,1}$, which is represented by the south pole of the left sphere. The first pulse in the sequence $S$ rotates this Bloch vector around the $x$-axis to a point in the $yz$-plane. The second pulse implements a $\pi$-rotation around the $y$-axis, flipping the $z$-component of the Bloch vector. Finally, the third pulse in $S$ performs the same rotation as the first pulse, thus aligning the Bloch vector with the north pole of the Bloch sphere.
In Fig.~\ref{fig:traj} on the right, we show the trajectories of an atom pair initially in the state $\ket{1,1}$, which is represented by the south pole of the right sphere. The first pulse in the sequence $S$ realizes a $\pi/2$ rotation around the $x$-axis on the Bloch sphere, aligning the Bloch vector with the $y$-axis. The second pulse corresponds to a rotation around the $y$-axis and thus leaves this state invariant. Finally, the third pulse in $S$ performs again a $\pi/2$ rotation around the $x$-axis, and thus rotates the Bloch vector to the north pole of the right Bloch sphere.  
By the same arguments, the second application of $S$ results in a return of the Bloch vector from the north to the south pole in all cases, i.e., $S\ket{0,r}=i\ket{0,1}$, $S\ket{r,0}=i\ket{1,0}$ and $S\ket{W}=i\ket{1,1}$. In summary, at the end of the protocol \eqref{eq:variant1} all computational basis states are mapped to themselves. In this process each of the computational basis states picks up a phase, which determines the gate performed in the qubit subspace. These phases can be directly inferred from the Bloch sphere picture, since they are given by half of the solid angle enclosed by the trajectories. For both trajectories in Fig.~\ref{fig:traj}, this solid angle is simply half of the area of the unit sphere, corresponding to a phase of $\pi$. Thus, all computational basis states except $\ket{0,0}$ acquire a phase of $\pi$ in the above process. This corresponds to the gate  $\ket{z_1,z_2}\rightarrow (-1)^{z_1 z_2-z_1-z_2}\ket{z_1,z_2}$, which is indeed the desired controlled-phase gate, CZ.

\begin{figure}[b]
    \centering
    \includegraphics[width=\linewidth]{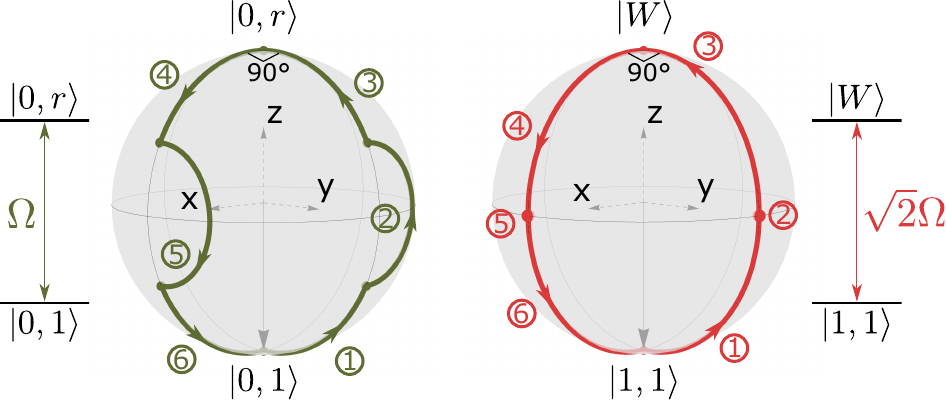}
    \caption{Bloch sphere trajectories for the controlled-($\tfrac{\pi}{2}$) gate sequence. The phase jump between the third and fourth pulses ensures that the area enclosed in the trajectories now represents one quarter of the sphere, so at the end of the sequence, the states $\ket{0,1}$, $\ket{1,0}$ and $\ket{1,1}$ have acquired a phase of $-\frac{\pi}{2}$. Again, $\ket{0,0}$ does not evolve, resulting in the desired gate.}
    \label{fig:piovertwogate}
\end{figure}

This 6-pulse sequence can easily be modified to realize controlled-phase gates with an arbitrary phase $\phi$, defined here as, 
\begin{align} \label{eq:Cphi}
    \textrm{C}_{\phi}\ket{z_1,z_2}= (e^{i\phi})^{z_1z_2-z_1-z_2}\ket{z_1,z_2}.
\end{align}
For this, one simply introduces a change of the laser phase by an amount $\xi$ between the first 3 and the last 3 pulses, i.e. between the first and second application of $S$. Formally this corresponds to shifting the value $\varphi\rightarrow\varphi+\xi$ in the last three pulses. In the Bloch sphere picture this  modifies the solid angle enclosed by each trajectory, and thus the phase acquired by the states $\ket{0,1}$, $\ket{1,0}$ and $\ket{1,1}$ (see Fig.~\ref{fig:piovertwogate}). This allows us to build any controlled-phase gate \eqref{eq:Cphi}, with the phase of the gate being determined by the phase jump via $\phi=\pi-\xi$ (see Appendix~\ref{App:arbphase} for details). A particularly useful case is the $\textrm{C}_{\pi/2}$ gate, where the corresponding pulse sequence is
\begin{align} \label{eq:Cpiovertwo}
\textrm{C}_{\pi/2}=U_y\lr{\alpha_1}U_{-x}\lr{\alpha_2}U_y\lr{\alpha_1}U_x\lr{\alpha_1}U_y\lr{\alpha_2}U_x\lr{\alpha_1}.
\end{align}
Analogous to the CZ case, two variants of the sequence exist, corresponding to the choices $\alpha_1=\tfrac{\pi}{2\sqrt{2}}$ and $\alpha_2=\pi$, or $\alpha_1=\tfrac{\pi}{2}$ and $\alpha_2=\tfrac{\pi}{\sqrt{2}}$, respectively. The Bloch sphere trajectories for the former choice are shown in Fig.~\ref{fig:piovertwogate}. 

Below we consider properties of the above protocols in the presence of noise in some control parameters. Since all the protocols defined in this section have the same properties in this regard, we refer to them indistinctly as Protocol \textrm{I} in the following.

\section{Robustness Analysis and robust gates} \label{Sec:robustness}

\begin{table*}[t]
\centering
\begin{tabular}{ |c|c|c|c|c|c| }
 \hline
 CZ gate protocol & Global drive & $\mc{F}$ & $\mc{P}$ & $\mc{C}$ & Execution time \\ 
 \hline
 Jaksch et al. \cite{jakschFastQuantumGates2000b} & {No} & {\color{black}$1-4.935\epsilon^2+O\lr{\epsilon^4}$} & {\color{black}$1-4.935\epsilon^2+O\lr{\epsilon^4}$} & {\color{Green}$1-4.870\epsilon^4+O\lr{\epsilon^6}$} & $12.57/\Omega$ \\ 
 Levine et al. \cite{levineParallelImplementationHighFidelity2019} & {Yes} & {\color{black}$1-2.963\epsilon^2+O\lr{\epsilon^3}$} & {\color{black}$1-2.547\epsilon^2+O\lr{\epsilon^3}$} & {\color{black}$1-0.416\epsilon^2+O\lr{\epsilon^3}$} & $8.59/\Omega$ \\ 
 \hline
 Protocol \textrm{I} & {Yes} & {\color{black}$1-1.878 \epsilon^2+O\lr{\epsilon^3}$} & {\color{black}$1-1.878 \epsilon^2+O\lr{\epsilon^3}$} & {\color{Green}$1-0.329\epsilon^4+O\lr{\epsilon^5}$} & $10.73/\Omega$ \\
 Protocol \textrm{II}  & {Yes} & {\color{Green}$1-0.329\epsilon^4+O\lr{\epsilon^5}$} & {\color{Green}$1-1.944\epsilon^6+O\lr{\epsilon^7}$} & {\color{Green}$1-0.329\epsilon^4+O\lr{\epsilon^5}$} & $21.45/\Omega$ \\
 \hline
\end{tabular}
\caption{Leading-order expansions for $\mc{F}$, $\mc{P}$ and $\mc{C}$ as a function of the laser intensity error parameter $\epsilon=\frac{\delta\Omega}{\Omega}$, for the protocols of \cite{jakschFastQuantumGates2000b} and \cite{levineParallelImplementationHighFidelity2019} and those presented here. We highlighted a vanishing susceptibility in green. Analytical expressions for all susceptibilities can be found in Appendix~\ref{App:tableanalytic}.}
\label{tab:tabfids}
\end{table*}

We note that Protocol \textrm{I} results in a total gate time $T_{\textrm{I}}=\frac{\pi}{\Omega}(2+\sqrt{2})$. While this is shorter than the protocol introduced in the seminal work by Jaksch et al.~\cite{jakschFastQuantumGates2000b}, it is slightly longer than the protocol recently implemented  in Ref.~\cite{levineParallelImplementationHighFidelity2019}. However, it has several advantages. All pulses are resonant, and they require only simple pulse times and phase jumps, potentially simplifying calibration in experiments. More importantly, all variants possess a native robustness against shot-to-shot fluctuations of the laser intensity, as we detail below.

To quantify such robustness, we consider several gate fidelity measures that are relevant in different scenarios below. For a given gate protocol,
we define a continuous family of unitary operators $U(\epsilon)$, parametrized by $\epsilon$, which measures the deviation of a control parameter (such as the Rabi frequency) from its target value. In particular, $U(0)$ is the target unitary operation realizing a controlled-phase gate in the qubit subspace.
The fidelity of $U(\epsilon)$ with the target gate is given by \cite{Pedersen_2007}
\begin{align}
\mc{F}=\frac{\mathrm{tr}\lr{PU(\epsilon)PU^\dag(\epsilon)}+|\mathrm{tr}\lr{U(\epsilon)PU^\dag({0}) P}|^2}{d(d+1)},
\end{align}
where $d=4$ is the dimension of the qubit subspace, and $P=\sum_{z_1,z_2\in\{0,1\}}\ket{z_1,z_2}\bra{z_1,z_2}$ is the projector onto it.
The first derivative of the fidelity vanishes at $\epsilon=0$ since $\mc{F}$ assumes its maximum there. Thus the fidelity susceptibility
\begin{align}
\chi=\frac{d^2}{d\epsilon^2}\mc{F}\big|_{\epsilon=0}
\end{align}
is a natural measure of the sensitivity of the protocol against variations of the corresponding control parameter \cite{shapiraRobustIon}. Accordingly, we call a protocol fully robust if $\chi=0$.  
Another important quantity when analyzing imperfect implementations of gate protocols is the average probability for the system to return to the qubit subspace at the end of the protocol,
\begin{align}
\mc{P}=\frac{\tr{PU(\epsilon)PU^\dag(\epsilon)}}{d}.
\end{align}
Clearly, at $\epsilon=0$ the leakage out of the qubit manifold vanishes and $\mc{P}=1$. Accordingly, we define a susceptibility 
\begin{align}
\chi_{\mc{P}}=\frac{d^2}{d\epsilon^2}\mc{P}\big|_{\epsilon=0}.
\end{align}
We call a protocol leakage-robust if $\chi_{\mc{P}}=0$.
Finally, we consider the conditional gate fidelity $\mc{C}$, defined as the fidelity of the gate, conditioned on the observation of no leakage out of the qubit subspace. It is defined as
\begin{align}
\mc{C}=\frac{1}{d+1}\lr{1+\frac{|\mathrm{tr}\lr{U(\epsilon)PU^\dag({0}) P}|^2}{\tr{PU(\epsilon)PU^\dag(\epsilon)}}}=\mc{F}/\mc{P}.
\end{align}
Leakage errors could be in principle detected and potentially converted to erasure errors, which can be corrected with a remarkably high threshold \cite{Wu2022NatComm}. The conditional fidelity $\mc{C}$ is thus particularly important, as it quantifies the effect of the remaining errors that can not be converted to erasure errors. Again, we define a corresponding susceptibility via 
\begin{align}
\chi_{\mc{C}}=\frac{d^2}{d\epsilon^2}\mc{C}\big|_{\epsilon=0}.
\end{align}
We call a gate conditionally robust if $\chi_\mc{C}=0$.
We have the relation $\chi=\chi_{\mc{P}}+\chi_{\mc{C}}$.

\subsection{Laser intensity errors}
\subsubsection{Conditional robustness}

In this section we are interested in the effect of variations of the laser intensity on gate fidelity. Experimentally these may originate from calibration errors or from drifts of the laser system that are slow on the timescale of the gate execution, i.e. $T_{\textrm{I}}$. To analyze these situations we set  $\epsilon\equiv\delta\Omega/\Omega$, where $\delta \Omega$ quantifies the deviation of the Rabi frequency from its target value. 
The interesting feature of Protocol \textrm{I} is that it is natively conditionally robust against such errors. It is straightforward to compute the leading order expansion of the conditional fidelity $\mc{C}$ for this type of error (see Appendix~\ref{App:phaseerror}), which reads 
\begin{multline}
    \mc{C}=1-\frac{\pi^4}{640}\left(13+4{\rm cos}\lr{\tfrac{\pi}{\sqrt{2}}}+8{\rm cos}\lr{\tfrac{2\pi}{\sqrt{2}}}-4{\rm cos}\lr{\tfrac{3\pi}{\sqrt{2}}}\right.\\
    \left.+3{\rm cos}\lr{\tfrac{4\pi}{\sqrt{2}}}\right)\epsilon^4+O\lr{\epsilon^5}.\nonumber
    \end{multline}
Indeed the second-order term vanishes identically, that is, $\chi_{\mc{C}}=0$ and the protocol is conditionally robust.

In Table \ref{tab:tabfids} we give the leading-order expansions for $\mc{F}$, $\mc{P}$ and $\mc{C}$ for Protocol \textrm{I}, as well as for the CZ gate protocols given in Refs. \cite{jakschFastQuantumGates2000b} and \cite{levineParallelImplementationHighFidelity2019}. Protocol \textrm{I} as well as the protocol of  Ref.~\cite{jakschFastQuantumGates2000b} are both conditionally robust, while the protocol of Ref.~\cite{levineParallelImplementationHighFidelity2019} is not. Note that the leading order contribution to the conditional fidelity is an order of magnitude smaller for Protocol \textrm{I}, compared to the protocol of Ref.~\cite{jakschFastQuantumGates2000b}. Moreover, we find that none of the three protocols are leakage-robust against laser intensity errors.

\subsubsection{Fully robust protocol}
To achieve full robustness ($\chi=0$), we now consider a variation of Protocol \textrm{I}. For this, the CZ gate is realized by applying two controlled-$(\frac{\pi}{2})$ gates in sequence, each of them realized according to Protocol \textrm{I}, with the pulse sequence \eqref{eq:Cpiovertwo}. 
Applying this sequence twice obviously realizes a CZ gate, with an execution time $T_{\textrm{II}}=2T_{\textrm{I}}$. The resulting protocol, which we call Protocol \textrm{II} from now on, inherits the conditional robustness of the individual $\textrm{C}_{\pi/2}$ sequences. Moreover, it is also leakage-robust against laser intensity errors. This is a result of the destructive interference between the two leakage amplitudes originating form imperfect implementations of each $\textrm{C}_{\pi/2}$ sequence. In fact it is straightforward to calculate the series expansion for $\mc{P}$,
    \begin{multline}
        \mc{P}=1-\frac{\pi^6}{1024}\left(30+30{\rm cos}\lr{\tfrac{\pi}{\sqrt{2}}}+27{\rm cos}\lr{\tfrac{2\pi}{\sqrt{2}}}\right.\\
    \left.+2{\rm cos}\lr{\tfrac{3\pi}{\sqrt{2}}}+6{\rm cos}\lr{\tfrac{4\pi}{\sqrt{2}}}+{\rm cos}\lr{\tfrac{6\pi}{\sqrt{2}}}\right)\epsilon^6+O\lr{\epsilon^7},\nonumber
    \end{multline}
which confirms leakage robustness, $\chi_\mc{P}=0$, but in addition shows that the contribution in next order also vanishes.
Since $\chi_{\mc{P}}=\chi_{\mc{C}}=0$, this Protocol II is fully robust to laser intensity errors. The leading-order expansions for ${\mc{F}}$, ${\mc{P}}$ and ${\mc{C}}$ in that case are presented in the fourth row of Table \ref{tab:tabfids}. We note that any controlled-phase gate $\textrm{C}_{\phi}$ \eqref{eq:Cphi} can similarly be realized in a fully robust way, by successively applying two $\textrm{C}_{\phi/2}$ sequences of the Protocol \textrm{I} type, separated by a well-chosen phase jump to ensure destructive interference of the leakage amplitudes (see Appendix~\ref{App:leakrob}).

\subsection{Motional Doppler shift} \label{Sec:Doppler}
Another source of gate errors in experiments with Rydberg atom arrays is thermal motion of the atoms in the traps \cite{grahamRydbergMediatedEntanglementTwoDimensional2019}. Specifically, the corresponding Doppler shift leads to a shot-to-shot fluctuation of the effective laser detuning. Since the velocities of the two atoms are uncorrelated, the corresponding Doppler shifts for the first atom, $\delta\Delta_1$, and for the second atom, $\delta\Delta_2$, break the permutation symmetry of the model by adding a term $-\sum_{i=1,2}\delta\Delta_{i}\ket{r}_{i}\bra{r}_{i}$ to the Hamiltonian \eqref{eq:Hamiltonian}. To analyze the robustness properties of the above gate protocols we find it convenient to work with the two independent uncorrelated parameters: $\frac{\delta\Delta_1+\delta\Delta_2}{2\Omega}$, which we call symmetric detuning error, and $\frac{\delta\Delta_1-\delta\Delta_2}{2\Omega}$, which we call antisymmetric detuning error. We note that robustness against errors of multiple parameters is equivalent to robustness against errors in each parameter individually  (see Appendix~\ref{App:multvar}). Note that this holds also for leakage-robustness and conditional robustness.
As a consequence, for a protocol to be robust against Doppler errors arising from the atoms' thermal motion, it must be both robust to  symmetric and anti-symmetric detuning errors individually. We therefore analyze these two cases separately below.

 \begin{figure*}[t]
    \centering
    \includegraphics[width=\textwidth]{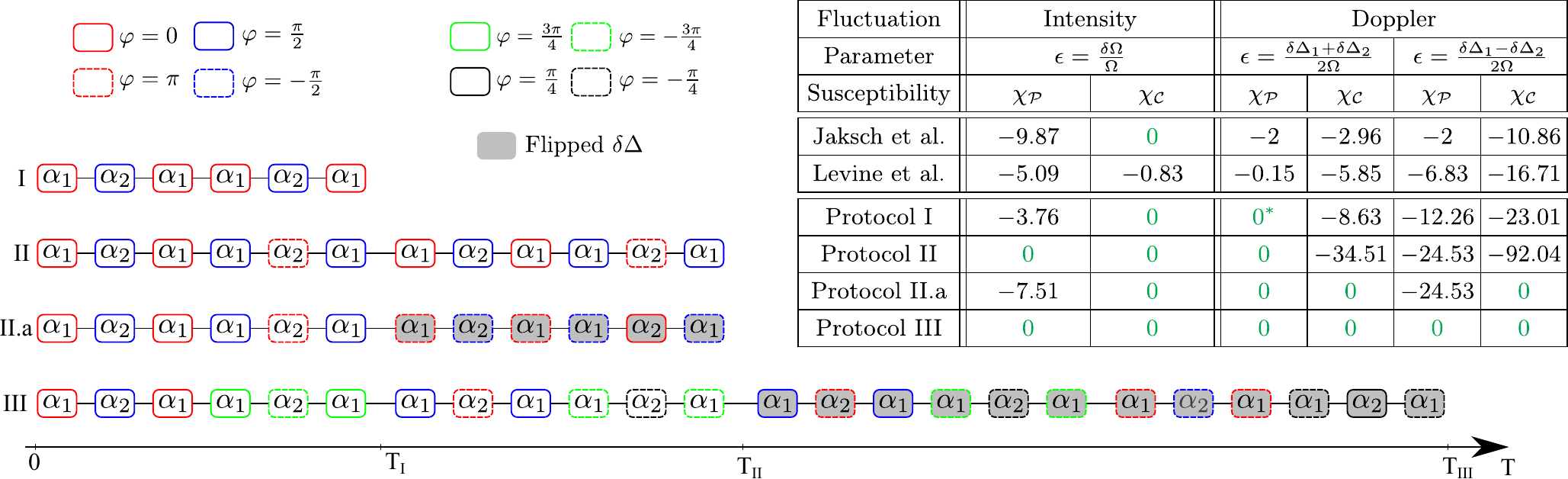}
    \caption{Left: Pulse sequences corresponding to the CZ gate protocols introduced here. Each global pulse is represented by one box, with the corresponding pulse area $\alpha_1$ or $\alpha_2$. The values of ($\alpha_1$, $\alpha_2$) can be chosen from two options: $(\tfrac{\pi}{2\sqrt{2}},\pi)$, as in \eqref{eq:variant1}, or $(\tfrac{\pi}{2},\tfrac{\pi}{\sqrt{2}})$, as in \eqref{eq:variant2}. The laser phase for each pulse is represented by the shape and color of the corresponding box's border. A filled box means the detuning error's sign is flipped for this pulse. The protocol duration is proportional to the number of boxes, and is indicated at the bottom of the figure.
    Right: Values of the leakage ($\chi_{\mc{P}}$) and conditional ($\chi_{\mc{C}}$) susceptibilities, for the error parameters associated to intensity or Doppler error, for all the protocols introduced above as well as the protocols of \cite{jakschFastQuantumGates2000b} and \cite{levineParallelImplementationHighFidelity2019}. The values given here correspond to a CZ gate with the choice $(\alpha_1,\alpha_2)=(\tfrac{\pi}{2\sqrt{2}},\pi)$; for an arbitrary phase and/or with $(\alpha_1,\alpha_2)=(\tfrac{\pi}{2},\tfrac{\pi}{\sqrt{2}})$, the specific values can be different but the robustness properties are the same (except for one case, denoted by a star in the table, where leakage-robustness is only achieved in the CZ case).}
    \label{fig:sequences}
\end{figure*}
 
\subsubsection{Symmetric detuning errors}

A symmetric detuning error corresponds to $\delta\Delta_1=\delta\Delta_2\equiv\delta\Delta$, so we define the error parameter $\epsilon=\frac{\delta\Delta}{\Omega}$. For symmetric detuning errors the permutaion symmetry is not broken and the dynamics can still be understood in terms of the effective two-level systems with corresponding Bloch spheres introduced above. A symmetric detuning error simply modifies the Bloch sphere trajectories: it tilts the rotation axes and changes the rotation angles for each of the pulses,  resulting in general in trajectories that do not close and thus in a reduced gate fidelity. Nevertheless, one can show analytically that the implementations of the CZ gate according to Protocol \textrm{I}  as well as realizations of arbitrary  C$_\phi$ gates according to Protocol II are natively leakage-robust against these symmetric Doppler errors, $\chi_{\mc{P}}=0$. This is in contrast to the protocols of Refs.~\cite{jakschFastQuantumGates2000b} and \cite{levineParallelImplementationHighFidelity2019}, which do not enjoy this feature.  
Unfortunately, neither Protocol I nor II are conditionally robust against symmetric detuning errors. 

To address this issue, we introduce an experimental method to invert the atomic velocities and thus echo out the Doppler shift. The central idea is to recapture the atoms in their harmonic optical tweezer potentials (which are otherwise turned off during the Rydberg pulses). After half a trap period, the velocity of each atom is reversed, and the tweezers turned off again. This effectively inverts the Doppler shift in the subsequent Rydberg pulses. We note that in this process not only the atomic velocity, but also the atomic position changes sign (when measured with respect to the trap minimum). 
An alternative to the above method to invert the Doppler shift is to invert the direction of the Rydberg laser (experimentally requiring calibration and stability of the relative path lengths of the two different paths for the laser). 
We note that in both these methods, fluctuations in the atomic position introduce an uncontrolled shift of the Rydberg laser phase between the pulses before and after this Doppler inversion. To avoid additional errors stemming from this effect, we only consider inverting the Doppler shifts at instances during a gate protocol, where all atomic populations are in qubit states, i.e., after the application of full gate sequences \footnote{This works as long as the uncontrolled phase shifts acquired in the Doppler inversion are small, as in typical experimental setups (Appendix~\ref{App:DopplerEcho}). If they are large, then the Doppler inversion should only be applied after the application of leakage robust sequences.}.

This possibility to invert the detuning error can be used to create protocols that are fully robust against this symmetric detuning error. To do so, we construct the $\textrm{CZ}$ gate out of two successive $\textrm{C}_{\pi/2}$ sequences, both realized according to Protocol \textrm{I}. We flip the sign of the detuning error between the first and the second C$_{\pi/2}$ gates, in order to echo out the relative phase error up to leading order and thus achieve conditional robustness. Additionally, we introduce a phase jump of $\pi$ between the two sequences, in order to ensure destructive interference of the Rydberg amplitudes, and thus leakage-robustness. The resulting protocol is thus fully robust. It is similar to Protocol \textrm{II} up to the inversion of the error's sign and the phase jump of $\pi$ between the two $\textrm{C}_{\pi/2}$ gates: we hence refer to it as Protocol \textrm{II}.a (see Fig.~\ref{fig:sequences}).
Once again, this protocol can easily be adapted to arbitrary controlled-phase gates. A fully robust  $\textrm{C}_{\phi}$ gate is realized by applying two successive $\textrm{C}_{\phi/2}$ gates (each following Protocol \textrm{I}): with an appropriate choice of the laser phase as well as an inverted detuning error for the second gate, the resulting gate sequence is fully robust against symmetric detuning errors.

The values of the leakage and conditional susceptibilities for symmetric detuning errors for the various protocols are given in Fig.~\ref{fig:sequences}. 
The full leading-order expansions of $\mc{F}$, $\mc{P}$ and $\mc{C}$ are given in Appendix~\ref{App:tableanalytic}.

\subsubsection{Antisymmetric detuning errors}

Now we consider antisymmetric variations of the frequency, which corresponds to opposite detuning errors for the two atoms, i.e., $\frac{\delta\Delta_1}{\Omega}=-\frac{\delta\Delta_2}{\Omega}\equiv\epsilon$. We first note that conditional robustness of a protocol against symmetric detuning errors implies that the same protocol also is conditionally robust against antisymmetric detuning errors. 
This is trivially true for the state $\ket{0,0}$, and can also be easily seen for the states $\ket{0,1}$ and $\ket{1,0}$: if the atoms are prepared in either of these two states, the effect of a symmetric and an anti-symmetric detuning error are the same, since only one of the atoms couples to the laser. Thus, for these states, the robustness of the dynamics against symmetric detuning errors directly implies robustness against antisymmetric errors as well.
The same is true also for the remaining state \ket{1,1}, but the analysis requires more care. In the presence of anti-symmetric detuning errors, the dynamics can no longer be reduced to an effective two-level system, but instead three states must be considered: the perturbation leads to an effective coupling of strength $\Omega\epsilon$ between $\ket{W}$ and the singlet state $\ket{A}=\frac{\ket{r,1}-\ket{1,r}}{2}$. Specifically, the dynamics of this three-level system is described by the Hamiltonian
\begin{align} \label{eq:3lshamilt}
    H_3=\Omega\lr{\frac{\sqrt{2}}{2} e^{i\varphi}\ket{1,1}\bra{W}-\epsilon\ket{W}\bra{A}}+\textrm{h.c.}
\end{align}
The form of this Hamiltonian directly implies that the diagonal matrix element of the corresponding evolution operator, $\bra{1,1}U(\epsilon)\ket{1,1}$, has no contribution in first order in $\epsilon$, which in turn results in a conditional robustness of the gate protocol against anti-symmetric detuning errors. 

We now discuss  leakage-robustness against anti-symmetric detuning errors. Again, for $\ket{0,0}$, $\ket{1,0}$ and $\ket{0,1}$, protocols that are leakage-robust against symmetric errors imply the same for anti-symmetric errors. For $\ket{1,1}$ the dominant leakage error stems form population in the $\ket{A}$ state, since population in the $\ket{W}$ state is of higher order in $\epsilon$. This can be seen from a penetrative analysis of the Hamiltonian \eqref{eq:3lshamilt}. To obtain a leakage-robust pulse sequence the final population in $\ket{A}$ must thus be cancelled up to second-order. This requirement can be expressed as $\bra{A}\partial_{\epsilon} U(0)\ket{1,1}=0$.

This can be satisfied by a new type of sequence, which we call Protocol \textrm{III} (see Fig.~\ref{fig:sequences}). In the same spirit as Protocols \textrm{II} and \textrm{II}.a, this variant consists in splitting the CZ gate into two $\textrm{C}_{\pi/2}$ gates. However, each of these gates is now realized according to Protocol \textrm{II}, that is, using a sequence of 12 pulses. We invert the detuning error (see the previous subsection) of the second $\textrm{C}_{\pi/2}$ sequence, which ensures conditional robustness of the CZ gate against symmetric and antisymmetric Doppler errors. Furthermore, we introduce a phase difference of $\tfrac{\pi}{2}$ between the first and second $\textrm{C}_{\pi/2}$ gates, which results in a destructive interference of all error terms, ensuring  $\bra{A}\partial_{\epsilon} U(0)\ket{1,1}=0$.
Finally, since both $\textrm{C}_{\pi/2}$ gates are realized with Protocol \textrm{II}, they are leakage-robust against symmetric frequency fluctuations, so the states $\ket{1,0}$ and $\ket{0,1}$ undergo robust dynamics as well. Thus, this new protocol, whose total duration is now $T_{\textrm{III}}=4T_{\textrm{I}}=42.90/\Omega$, is fully robust to symmetric and anti-symmetric Doppler frequency errors. The pulse sequence and associated susceptibilities are shown in Fig.~\ref{fig:sequences}.

The protocols \textrm{II}.a and \textrm{III} have the added advantage of having simultaneous (conditional or full) robustness against intensity and Doppler errors. Indeed, so far we have discussed robustness of gate protocols against intensity and Doppler errors separately, but in practice it would be desirable to identify protocols that are simultaneously robust against both. The table in Fig.~\ref{fig:sequences} gives us the corresponding information: we see that Protocols \textrm{II}.a and \textrm{III} are the most interesting in this context. Indeed, the former is conditionally robust against variations of all three error parameters corresponding to intensity or Doppler errors, and the latter is fully robust against them all (at the cost of a longer execution time).

Finally, we note that the required pulse time can be compressed if we relax the requirement that Doppler inversion be performed between full gates. This can be interesting in the limit where the atoms are tightly trapped, such that fluctuations of the atoms' positions are very small relative to the optical wavelength, and thus the error due to the positional phase shift is negligible compared to other error sources. We introduce two additional protocols in Appendix~\ref{App:protIaIIb} where the Doppler shift inversion is realized while the qubits are in the Rydberg state, that result in conditionally- or fully-robust dynamics against intensity and Doppler fluctuations with shorter exectution times.

\section{Conclusion}
We have introduced novel protocols to realize arbitrary controlled-phase gates between two neutral atoms based on resonant global excitation to Rydberg states and the Rydberg blockade mechanism. A main feature of these protocols is that they exhibit robustness against certain quasi-static coherent errors. Our analysis focused on effects due to errors in calibration of the Rabi frequency or errors arising from slow fluctuations of the laser intensity, as well as errors from Doppler shifts. We note that in current experimental setups, other error sources, such as spontaneous emission from Rydberg states or from intermediate states that are virtually populated during the Rydberg excitation process, often dominate the error budget and a detailed analysis of each setup is necessary to assess whether or not these protocols can contribute to improved gate fidelity in practice. Finally we note that the ideas introduced here for two-qubit gates also generalize to a multi-qubit setting (see Appendix~\ref{Sec:3qubit}).

\section*{Acknowledgements}
We thank Manuel Endres, Mikhail Lukin, and Nishad Maskara for helpful discussions. We acknowledge financial support from the ERC Starting grant QARA (grant no.~101041435), and an ESQ Discovery Grant, the Army Research Office (grant no.~W911NF-21-1-0367) and the DARPA ONISQ program (grant no.~W911NF2010021).  D.B. acknowledges support from the NSF Graduate Research Fellowship Program (grant DGE1745303) and The Fannie and John Hertz Foundation. 

\textit{Note added} --- While finalizing this work we became aware of Ref.~\cite{janduraOptimizingGates}. 

\section*{Appendix}
\appendix

\section{Arbitrary controlled-phase gate} \label{App:arbphase}

Here we detail how to use Protocol \textrm{I} to realize a controlled-phase gate with arbitrary phase $\phi$ (see Eq.~\eqref{eq:Cphi}). The idea is to use the same pusle sequence as in eq.~\eqref{eq:variant2} (interpreted as a 6-pulse sequence as discussed in the main text), but change the phase of the laser in the last three pulses by an amount $\xi$. The full evolution unitary is thus (we use here the sequence \eqref{eq:variant2} as our basis, but the analysis is the same with \eqref{eq:variant1})
\begin{align}
    U=U_{x,\xi}(\tfrac{\pi}{2})U_{y,\xi}(\tfrac{\pi}{\sqrt{2}})U_{x,\xi}(\tfrac{\pi}{2})U_x(\tfrac{\pi}{2})U_y(\tfrac{\pi}{\sqrt{2}})U_x(\tfrac{\pi}{2}),
\end{align}
where the subscript $\xi$ indicates that the phase is shifted by $\xi$. Defining
\[P_{\xi}=\prod_{i=1,2} e^{i\xi}\ket{r}_i\bra{r}_i+\ket{1}_i\bra{1}_i+\ket{0}_i\bra{0}_i,\]
one can write $U_{x,\xi}(\theta)=P_{\xi}^{\dagger}U_x(\theta)P_{\xi}$. Using the fact that $P_{\xi}^{\dagger}P_{\xi}=\mathbb{1}$, this leads to
\begin{align}
    U=P_{\xi}^{\dagger}S \, P_{\xi}S,
\end{align}
where $S=U_x(\tfrac{\pi}{2})U_y(\tfrac{\pi}{\sqrt{2}})U_x(\tfrac{\pi}{2})$.

As seen in the main text, $S$ is designed so that $S\ket{1,0}\propto\ket{r,0}$, $S\ket{0,1}\propto\ket{0,r}$ and $S\ket{1,1}\propto\frac{\ket{r,1}+\ket{1,r}}{\sqrt{2}}$. All these states have exactly one atom in the Rydberg state, so for $\ket{z_1,z_2}\neq\ket{0,0}$, we have
\begin{align}
    P_{\xi}S\ket{z_1,z_2}=e^{i\xi}S\ket{z_1,z_2}.
\end{align}
Therefore, 
\begin{align}
    U\ket{z_1,z_2}=P_{\xi}^{\dagger}S\,P_{\xi}S\ket{z_1,z_2}=e^{-i\lr{\pi-\xi}}\ket{z_1,z_2}.
\end{align}
Finally, since $U\ket{0,0}=\ket{0,0}$, it follows that for any $\ket{z_1,z_2}$,
\begin{align}
    U\ket{z_1,z_2}=\lr{e^{i(\pi-\xi)}}^{z_1z_2-z_1-z_2}\ket{z_1,z_2},
\end{align}
which is a controlled-$\phi$ gate with $\phi=(\pi-\xi)$, up to a single-qubit rotation. In summary, to realize a controlled-phase gate of phase $\phi$, one needs to separate both halves of the 6-pulse sequence by a phase jump $\xi=\pi-\phi$.

\section{Extension to more than two qubits} \label{Sec:3qubit}

For practical applications and the implementation of quantum circuits, it is useful to be able to realize entangling gates for more than two qubits. Indeed, although all multiqubit gates can be decomposed as a product of one- and two-qubit gates, the number of gates required scales quickly with the number of qubits, and can make such decompositions impractical. In this section, we therefore discuss the extension of the protocols presented above to more than two atoms.
We now consider three atoms that are all in the same blockade radius: as a result, no more than one of them may be excited to the Rydberg state at a time. Same as above, the dynamics thus decomposes as a block-diagonal evolution, but there is now a third effective two-level system to consider, corresponding to the state $\ket{1,1,1}$ and with effective Rabi frequency $\sqrt{3}\Omega$.

Our main target is now the Controlled-Controlled-Z gate (CCZ), which is defined as ${\rm{CCZ}}\ket{z_1,z_2,z_3}=(-1)^{z_1z_2z_3}\ket{z_1,z_2,z_3}$. We numerically find a sequence $S_3$ of five resonant global pulses, which brings all three two-level systems from their ground state to their excited state: writing a pulse with pulse area $\alpha=\Omega t$ and phase $\xi$ as $U_{\xi}(\alpha)$, the five-pulse sequence $S_3$ reads
\begin{align}
    S_3=U_{0}(\alpha_1)U_{\xi_2}(\alpha_2)U_{\xi_3}(\alpha_3)U_{\xi_2}(\alpha_2)U_{0}(\alpha_1)
\end{align}
where $\alpha_1=1.088$, $\alpha_2=1.955$, $\alpha_3=5.373$, $\xi_2=1.552$, and $\xi_3=1.593$. Just like in Sec.~\ref{Sec:gate}, applying this sequence twice brings all computational basis states back to themselves, and gives them a minus sign (except $\ket{0,0,0}$). This is equivalent to the above definition of the CCZ gate up to exchanging the states $\ket{0}$ and $\ket{1}$. The total execution time for this protocol is $T=22.84/\Omega$, which is only about twice as long as the two-qubit protocol introduced in the main text. Moreover, this protocol can also be adapted to realize arbitrary Controlled-Controlled-Phase gates ($\textrm{CC}_{\phi}$), by introducing a phase jump between the two successive applications of $S_3$.

\begin{figure}[ht]
    \centering
    \includegraphics[width=\linewidth]{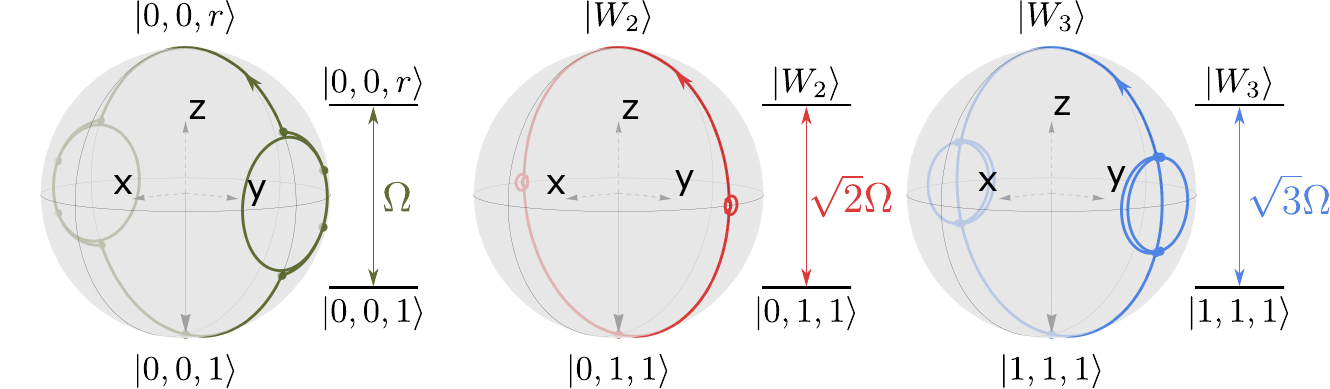}
    \caption{Bloch sphere representation of the 3-qubit CCZ gate sequence. The coupled superposition states $\ket{W_2}$ and $\ket{W_3}$ are defined respectively as $\ket{W_2}=\frac{\ket{0,1,r}+\ket{0,r,1}}{\sqrt{2}}$ and $\ket{W_3}=\frac{\ket{1,1,r}+\ket{1,r,1}+\ket{r,1,1}}{\sqrt{3}}$; the associated coupling strengths are respectively $\sqrt{2}\Omega$ and $\sqrt{3}\Omega$. For all three Bloch spheres, the solid angle enclosed in the trajectory represents half the sphere, so all computational basis states except $\ket{0,0,0}$ acquire a phase of $\pi$ during this sequence.}
    \label{fig:CCZgate}
\end{figure}
Finally, the protocol introduced here retains the robustness properties of its two-qubit counterpart against laser intensity fluctuations. Indeed, the division of the sequence in two applications of $S_3$,  which maps the south poles of all three Bloch spheres to their north poles, ensures conditional robustness for the same reasons as those given in Appendix~\ref{App:phaseerror}. Moreover, in the same way as in Sec.~\ref{Sec:robustness}, dividing a $\textrm{CC}_{\phi}$ gate into two successive $\textrm{CC}_{\phi/2}$ gates, separated by a well-chosen phase jump (which is zero in the CCZ case), gives rise to a destructive interference which ensures leakage-robustness of the gate at the cost of a doubled execution time.

Extension of these principles to more than three qubits is similarly possible, but the numerical cost of finding a sequence that brings all effective two-level systems from their ground to excited state increases with the number of atoms considered. Still, once such a sequence is found, the robustness ideas developed in Sec.~\ref{Sec:robustness} apply to the resulting gate protocol as well.

\section{Conditional robustness of Protocol \textrm{I}} \label{App:phaseerror}

\subsubsection{Requirement for conditional robustness}
In this section we show  that Protocol \textrm{I} is conditionally robust against intensity fluctuations. For this we first note that a gate is conditionally robust if 
$\bra{z_1,z_2}\partial_{\epsilon}U(0)\ket{z_1,z_2}=0$ for $(z_1,z_2)\in\{(0,1),(1,0),(1,1)\}$, where $\partial_{\epsilon}U(0)\equiv \partial_\epsilon U(\epsilon)|_{\epsilon=0}$. Indeed, in that case we have (keeping in mind that $U(\epsilon)\ket{0,0}=\ket{0,0}$ for any $\epsilon$)
\begin{align*}
|\mathrm{tr}&\lr{U(\epsilon)PU^\dag({0}) P}|^2\\
&\qquad=|\sum_{\vec{z}}\bra{\vec{z}}U(\epsilon)\ket{\vec{z}}\bra{\vec{z}}U(0)\ket{\vec{z}}|^2\\
&\qquad=|4-\frac{\epsilon^2}{2}\sum_{\vec{z}\neq(0,0)}\bra{\vec{z}}\partial_{\epsilon}^2 U(0)\ket{\vec{z}}+O\lr{\epsilon^4}|^2\\
&\qquad=4\lr{4-\epsilon^2\sum_{z_1,z_2}Re\left[\bra{\vec{z}}\partial_{\epsilon}^2 U(0)\ket{\vec{z}}\right]}+O\lr{\epsilon^4},
\end{align*}
and 
\begin{align*}
\mathrm{tr}&\lr{PU(\epsilon)PU^\dag({\epsilon})}\\
&\qquad=1+\sum_{\vec{z}\neq(0,0)}|\bra{\vec{z}}U(\epsilon)\ket{\vec{z}}|^2\\
&\qquad=1+\sum_{\vec{z}\neq(0,0)}|1-\frac{\epsilon^2}{2}\bra{\vec{z}}\partial_{\epsilon}^2 U(0)\ket{\vec{z}}+O\lr{\epsilon^4}|^2\\
&\qquad=4-\epsilon^2\sum_{z_1,z_2}Re\left[\bra{\vec{z}}\partial_{\epsilon}^2 U(0)\ket{\vec{z}}\right]+O\lr{\epsilon^4},
\end{align*}
which, from the expressions of $\mc{F}$ and $\mc{P}$, leads to
\[\mc{C}=\frac{\mc{F}}{\mc{P}}=1-O\lr{\epsilon^4}.\]
This is indeed the definition of conditional robustness.

We note that the requirement $\bra{z_1,z_2}\partial_{\epsilon}U(0)\ket{z_1,z_2}=0$ corresponds to saying that there is no first-order phase error in the $\ket{z_1,z_2}$ component of the final state. Indeed, this matrix element gives the first-order term of the series expansion of $\bra{z_1,z_2}U(\epsilon)\ket{z_1,z_2}$, which is equal to $i$ times the first-order phase error.

\subsubsection{Absence of phase error in Protocol I}
We thus need to show that $\bra{z_1,z_2}\partial_{\epsilon}U(0)\ket{z_1,z_2}=0$ for all qubit states. We write $U$ as (using the same notations as the previous section)
\begin{align*}
U(\epsilon)=P_{\xi}^{\dagger}S(\epsilon) \, P_{\xi}S(\epsilon),
\end{align*}
where $S(\epsilon)$ is defined analogous to $S$. Suppressing the explicit dependence on $\epsilon$ and introducting the notation $U'\equiv\partial_{\epsilon}U$ and $S'\equiv\partial_{\epsilon}S$ for simplicity, we can then write $\partial_{\epsilon}U$ as
\begin{align}
U'=P_{\xi}^{\dagger}S' \, P_{\xi}S+P_{\xi}^{\dagger}S \, P_{\xi}S'.
\end{align}
Defining $\ket{\uparrow_{z_1,z_2}}=S(0)\ket{z_1,z_2}$, we have \[S(0)\ket{\uparrow_{z_1,z_2}}=S(0)^2\ket{z_1,z_2}=-\ket{z_1,z_2},\]
hence
\[S^{\dagger}(0)\ket{z_1,z_2}=-\ket{\uparrow_{z_1,z_2}},\]
From this, and using the fact that (since $\ket{\uparrow_{z_1,z_2}}$ has exactly one atom in the Rydberg state) $P_{\xi}\ket{\uparrow_{z_1,z_2}}=e^{i\xi}\ket{\uparrow_{z_1,z_2}}$ and $\bra{\uparrow_{z_1,z_2}}P_{\xi}=e^{i\xi}\bra{\uparrow_{z_1,z_2}}$, we can write
\begin{align} \label{eq:vanishingphasecondition}
\begin{split}
\bra{z_1,z_2}U'\ket{z_1,z_2}&=e^{i\xi}\left(\bra{z_1,z_2}S'\ket{\uparrow_{z_1,z_2}}\right.\\
&\qquad\qquad\left.-\bra{\uparrow_{z_1,z_2}}S'\ket{z_1,z_2}\right)
\end{split}
\end{align}
We now take advantage of the block-diagonal evolution of the system to consider each computational basis state's evolution as an independent two-level system, with ground state $\ket{z_1,z_2}$ and excited state $\ket{\uparrow_{z_1,z_2}}$. 

We consider the restrictions of all 8-by-8 operators to the two-level system $\{\ket{z_1,z_2},\ket{\uparrow_{z_1,z_2}}\}$ from now on.

$S(\epsilon)$ is the product of three single-pulse evolution operators, $S(\epsilon)=\prod_{p=1}^{3}U_{p}(\epsilon)$ with $U_{p}=e^{-iH_{p}t_{p}}$, where the product goes over the three pulses of $S$ (implying an order in the product) and the subscript $p$ specifies which pulse each operator corresponds to. For the two-level system $\{\ket{z_1,z_2},\ket{\uparrow_{z_1,z_2}}\}$, each single-pulse unitary $U_{p}$ is a Bloch sphere rotation, so their product can also be expressed as one: the restriction of $S(\epsilon)$ to $\{\ket{z_1,z_2},\ket{\uparrow_{z_1,z_2}}\}$ is
\begin{align}
    S(\epsilon)&=e^{i\frac{\theta(\epsilon)}{2}\vec{n}(\epsilon)\cdot\vec{\sigma}}\\
    &=e^{i\lr{\frac{\theta(0)}{2}\vec{n}(0)\cdot\vec{\sigma}+\frac{\epsilon}{2}\partial_{\epsilon}\lr{\theta(\epsilon)\vec{n}(\epsilon)}\cdot\vec{\sigma}}+O\lr{\epsilon^2}},
\end{align}
where $\vec{\sigma}$ is the vector containing the three Pauli matrices of the two-level system $\{\ket{z_1,z_2},\ket{\uparrow_{z_1,z_2}}\}$. The commutator of two Pauli matrices is either proportional to another Pauli matrix or 0. Thus, one can rewrite this (using the Baker-Campbell-Hausdorff formula) as 
\begin{align}
    S(\epsilon)&=e^{i\epsilon\frac{\alpha}{2}\vec{m}\cdot\vec{\sigma}}S(0)+O\lr{\epsilon^2}
\end{align}
with some real scalar $\alpha$ and vector $\vec{m}$. From this one gets
\begin{align}
    S(\epsilon)&=S(0)+i\frac{\alpha}{2}\lr{\vec{m}\cdot\vec{\sigma}}S(0)\epsilon+O\lr{\epsilon^2}.
\end{align}
The derivative of $S$ is given by the second term of this expression.
Finally, this means that
\begin{align*}
\bra{z_1,z_2}S'\ket{\uparrow_{z_1,z_2}}
&=-i\frac{\alpha}{2}\bra{z_1,z_2}\vec{m}\cdot\vec{\sigma}\ket{z_1,z_2}\\
&=i\frac{\alpha}{2}\vec{m}\cdot\vec{e}_z
\end{align*}
and
\begin{align*}
\bra{\uparrow_{z_1,z_2}}S'\ket{z_1,z_2}&=i\frac{\alpha}{2}\bra{\uparrow_{z_1,z_2}}\vec{m}\cdot\vec{\sigma}\ket{\uparrow_{z_1,z_2}}\\
&=i\frac{\alpha}{2}\vec{m}\cdot\vec{e}_z.
\end{align*}
Plugging this into \eqref{eq:vanishingphasecondition} finally shows that, for all $\ket{z_1,z_2}$, we have
\[\bra{z_1,z_2}\partial_{\epsilon}U(0)\ket{z_1,z_2}=0,\]
which, as shown in the beginning of this section, implies that the protocol is conditionally robust.

\section{Leakage-robustness of Protocol \textrm{II}} \label{App:leakrob}
Here we show that Protocol \textrm{II} is leakage-robust against laser intensity errors, for an arbitrary controlled-phase gate.

The gate is realized by applying two successive controlled-phase gates of half the target phase, such that the resulting gate is the desired one.  One also needs to introduce a phase jump between the two successive gates. The pulse sequence can be written as
\begin{align}
    U(\epsilon)=\lr{P_{\xi}^{\dagger}G(\epsilon) P_{\xi}}G(\epsilon),
\end{align}
where $\phi$ is the target phase, $G$ the 6-pulse sequence corresponding to a controlled-$(\tfrac{\phi}{2})$ gate (see Sec.~\ref{Sec:gate} and Appendix~\ref{App:arbphase}), and $\xi$ the phase jump between the two gates. As in the previous section, we denote $\partial_{\epsilon}U$ as $U'$ and $\partial_{\epsilon}G$ as $G'$.

The sequence $G$ is designed such that $G(0)\ket{z_1,z_2}=e^{i\tfrac{\phi}{2}f(z_1,z_2)}\ket{z_1,z_2}$ where $f(z_1,z_2)=z_1 z_2-z_1-z_2$, and one can check that with $Q=\ket{r,0}\bra{r,0}+\ket{0,r}\bra{0,r}+\ket{W}\bra{W}$ the projector onto the manifold of coupled Rydberg states, $QG(0)=e^{i\tfrac{\phi}{2}}Q$, so at $\epsilon=0$,
\begin{align}
    QU'\ket{z_1,z_2}&=Q\left[\lr{P_{\xi}^{\dagger}G' P_{\xi}}G+\lr{P_{\xi}^{\dagger}G P_{\xi}}G'\right]\ket{z_1,z_2} \notag\\
    &=Q\left[e^{i\tfrac{\phi}{2}f(z_1,z_2)}e^{-i\xi}+e^{i\tfrac{\phi}{2}}\right]G'\ket{z_1,z_2},
\end{align}
where we used the fact that $QP_{\xi}^{\dagger}=e^{-i\xi}Q$ (since all the states in $Q$'s manifold have exactly one atom in the Rydberg state).
For $\ket{z_1,z_2}=\ket{0,0}$, this vanishes since $G'\ket{0,0}=0$, and for $\ket{z_1,z_2}\neq\ket{0,0}$, we have $f(z_1,z_2)=-1$, so 
\begin{align}
    QU'(0)\ket{z_1,z_2}&=Q\left[e^{-i\lr{\xi+\tfrac{\phi}{2}}}+e^{i\tfrac{\phi}{2}}\right]G'(0)\ket{z_1,z_2}
\end{align}
Thus, for $\xi=\pi-\phi$, we have $QU'(0)\ket{z_1,z_2}=0$. This leads to $QU(\epsilon)P=O\lr{\epsilon^2}$.

From this, since $(P+Q)U(\epsilon)=U(\epsilon)(P+Q)$, we can write
\begin{align*}
\mc{P}&=\frac{1}{d}\tr{PU(\epsilon)PU(\epsilon)^{\dagger}}\\
&=1-\frac{1}{d}\tr{PU(\epsilon)QU(\epsilon)^{\dagger}P}\\
&=1-O\lr{\epsilon^4}
\end{align*}
which means $\chi_{\mc{P}}=0$.

We note that for the CZ gate case, this analysis leads to a phase jump $\xi=\pi-\pi=0$, which explains why no phase jump is needed in the CZ case.

\section{Multivariate robustness} \label{App:multvar}
In practice, more than one control parameter may fluctuate. We prove here that a gate is robust to simultaneous variations of two or more parameters if it is robust with respect to each control parameter separately.

We replace the error parameter $\epsilon$ with a vector of control parameters $\vec{\epsilon}=(\epsilon_1,\epsilon_2,\dots)$. Since the fidelity assumes its maximum at $\vec{\epsilon}=\vec{0}$, its Hessian matrix  $H_{ij}=\partial_{\epsilon_i}\partial_{\epsilon_j} \mc{F}(\vec{\epsilon})|_{\vec{\epsilon}=\vec{0}}$, is negative semi-definite. In particular, the 2-by-2 submatrix $\big(\begin{smallmatrix}
  H_{ii} & H_{ij}\\
  H_{ji} & H_{jj}
\end{smallmatrix}\big)$ is also negative semi-definite, so its determinant must be positive, such that (as $H_{ij}=H_{ji}$)
\begin{align}
\begin{split}
&H_{ii}H_{jj}-H_{ij}^2\geq0\\
&(\partial_{\epsilon_i}\partial_{\epsilon_j} \mc{F})^2 \leq (\partial_{\epsilon_i}^2 \mc{F}) (\partial_{\epsilon_j}^2 \mc{F}).
\end{split}
\end{align}
If a protocol is robust to variations of the two parameters $\epsilon_i$ and $\epsilon_j$, the right hand-side is zero, so the left hand-side must also be zero, meaning that the sequence is also robust to simultaneous fluctuations.

\section{Short robust protocols with Doppler inversion in the Rydberg state}\label{App:protIaIIb}

\begin{figure}[t]\label{fig:seqProtIaIIb}
    \centering
    \includegraphics[width=\linewidth]{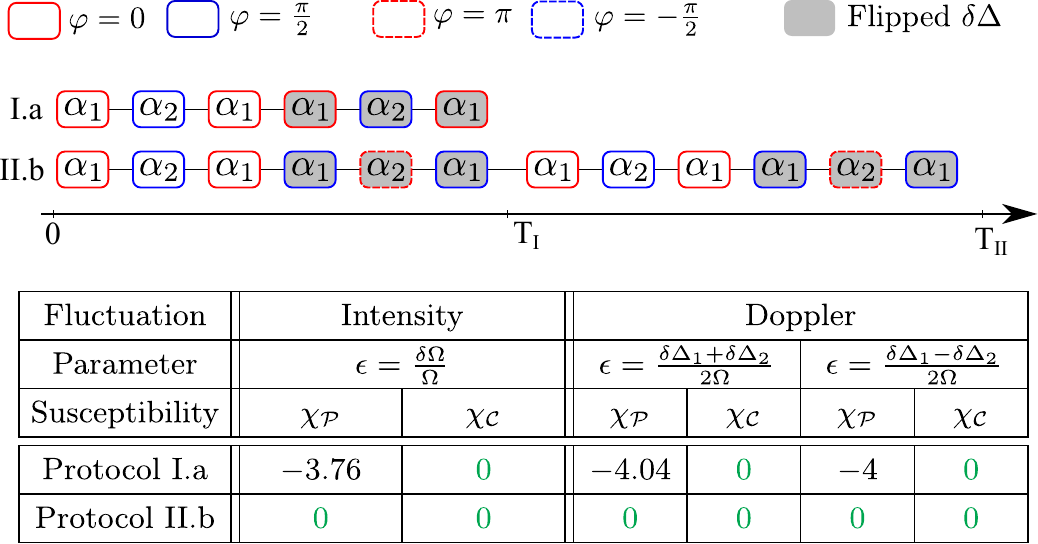}
    \caption{CZ pulse sequences for Protocols \textrm{I}.a ($T_{\textrm{I}}=10.73/\Omega$) and \textrm{II}.b ($T_{\textrm{II}}=21.45/\Omega$), and susceptibilities of these sequences against intensity and Doppler errors.}
    \label{fig:protIaIIb}
\end{figure}

We now introduce two new protocols that take advantage of tightly trapped atoms, i.e. we assume that the effects of positional phase shift due to the Doppler inversion scheme can be neglected. In that case, one can create a protocol that is conditionally robust to (symmetric and antisymmetric) Doppler errors using only 6 pulses. This protocol (which we call Protocol \textrm{I}.a) simply consists in realizing a 6-pulse Protocol \textrm{I}, but reversing the Doppler shift for the last three pulses (second half) of the sequence. Computing the expansions of the conditional susceptibilities in that case shows that the sequence is indeed conditionally robust to symmetric and antisymmetric Doppler errors. In addition, due to being the same as Protocol \textrm{I} up to the Doppler inversion, this protocol is also conditionally robust against laser intensity errors.

The second protocol we introduce is a fully robust variant of Protocol \textrm{I}.a. Its structure follows naturally from the same ideas as in the main text: we split the $\textrm{C}_{\phi}$ gate into two Protocol \textrm{I}.a-style $\textrm{C}_{\phi/2}$ gates, and separate them by the right phase difference (0 for the CZ gate) to ensure destructive interference of the leakage amplitudes. Computing the expansions of the fidelities in this case shows that this protocol (which we call Protocol \textrm{II}.b) is fully robust against intensity and Doppler errors. The CZ pulse sequences and associated susceptibilities for Protocols \textrm{I}.a and \textrm{II}.b are given in Fig.~\ref{fig:seqProtIaIIb}.

\section{Analysis of the Doppler echo scheme}\label{App:DopplerEcho}

\subsubsection{The scheme}
In this section, we further describe the Doppler echo scheme discussed in the text and provide further analysis. Each individual atom resides in its own individual optical tweezer, which acts as a harmonic oscillator potential. The atoms are typically not cooled all the way to the ground state of this quantum harmonic oscillator, and in particular are in a thermal state of the oscillator \cite{kaufmanCoolingSingleAtom}, which can be described as the atom being in a random coherent state with random amplitude and phase. The motional state of the atom at any point in time can thus be understood as simply oscillating in a classical harmonic oscillator potential with some random amplitude and phase of oscillation. The atom's position and velocity in the classical harmonic oscillator as a function of time will be

\begin{align*}
    x(t) = x_0 \cos(\omega_0 t + \phi) \\
    v(t) = v_0 \sin(\omega_0 t + \phi)
\end{align*}
where $\omega_0$ is the trap frequency and $x_0, v_0, \phi$ are random position, velocity, and phase. The Doppler error is proportional to the velocity $v(t)$ of the atom at the time that the gate is performed. The time of the gate (during which the tweezer is typically turned off \cite{levineParallelImplementationHighFidelity2019}) is significantly shorter than the trap frequency, and so the atom velocity and position in the oscillator does not significantly change during the gate. Upon recapturing the atoms in the tweezer and waiting half a trap period, i.e. $\pi / \omega_0$, the atom position and velocity then reverse:

\begin{align*}
    x(\pi / \omega_0) = -x(0) \\
    v(\pi / \omega_0) = -v(0)
\end{align*}

As such, the Doppler error $\propto v$ will reverse in sign, as will the positional fluctuation of the atom about the center of the potential. This can be used to echo and suppress Doppler-induced dephasing. With all population stored in the qubit states $\{ \ket{0}$, $\ket{1} \}$ after a gate, any Rydberg laser phase noise or dephasing between $\ket{1}$ and $\ket{r}$ during the half-trap-period oscillation will cause no effect since all phase information is stored between $\ket{0}$ and $\ket{1}$.

However, if one leaves population in the Rydberg state $\ket{r}$ during the half-trap-period oscillation, several complications arise. First, there can be significant decoherence between $\ket{1}$ and $\ket{r}$ from both inhomogeneous light shifts from the tweezer and from spin-motion entanglement induced by the different trapping potentials for $\ket{1}$ and $\ket{r}$ (in fact, $\ket{r}$ is anti-trapped for many settings \cite{bluvsteinQuantumProcessorBased2022_2}). To avoid such decoherence, one requires a magic trapping potential, i.e. such that both $\ket{1}$ and $\ket{r}$ experience the same trapping potential, which can be made possible in various settings \cite{zhangMagictrapping}.

Even with a magic trapping potential, dephasing can still occur between $\ket{1}$ and $\ket{r}$ if population is left in $\ket{r}$ for the half-trap-period oscillation. Specifically, the reversal of the atom position along the direction of the laser is in fact precisely a Doppler shift: the atom moves a distance $2 x(0)$ which will cause it to accumulate phase within the wavelength of the Rydberg laser by an amount $k \times 2 x(0)$, where $k$ is the net momentum of the Rydberg laser (for a single laser this is just $k = 2\pi/\lambda$ where $\lambda$ is the laser wavelength, and for a two-photon counter-propagating excitation scheme this is the net momentum between the two lasers $k_1 - k_2$). Since $x(0)$ is randomly drawn from the thermal distribution of the atom position, the phase accumulation $k \times 2 x(0)$ will now lead to a random phase fluctuation between $\ket{1}$ and $\ket{r}$. Quantitatively, the thermal spread of the atom can be attained from the equipartition theorem: for a massive particle $m$ the root-mean-square spread in position will be $x_{\text{RMS}} = \sqrt{k_B T / (m \omega_0^2)}$, where $k_B$ is Boltzmann's constant and $T$ is the atom temperature. For $^{87}$Rb, with typical trap frequencies on the scale of $\omega_0 \sim 2\pi \times 100$ kHz and atom temperatures on the scale of $\sim$ 5 $\mu$K \cite{bluvsteinQuantumProcessorBased2022_2,kaufmanCoolingSingleAtom}, this will give $x_{\text{RMS}} \sim $ 35 nm. For the two-photon scheme used for $^{87}$Rb \cite{bluvsteinQuantumProcessorBased2022_2, levineParallelImplementationHighFidelity2019}, this would lead to a characteristic phase accumulation of $k \times 2 x_{\text{RMS}} \sim 2\pi \times 0.1$ rad. This is a non-negligible phase accumulation. Thus, it is useful to determine the effect of such a fluctuation for the protocols discussed above, in the cases where it cannot be suppressed.

\subsubsection{Effect of the positional phase shift, by protocol}
The first thing to consider is whether this phase fluctuation, by itself, causes a gate infidelity. There are two different cases in the protocols discussed in Sec.~\ref{Sec:robustness} and Appendix~\ref{App:protIaIIb}: either the Doppler inversion occurs between two controlled-phase gates, while the atoms are in the qubit manifold (Protocol \textrm{II}.a and Protocol III), or it happens in the middle of a 6-pulse gate sequence, while they are in the Rydberg-excited state (Protocols \textrm{I}.a and \textrm{II}.b). In the first case, a different laser phase does not modify the action of the second $\textrm{C}_{\phi/2}$ gate: the two gates still add up normally and give the desired $\textrm{C}_{\phi}$ gate. The phase shift thus has no effect in that case. In the second case, however, the phase shift effectively modifies the phase jump between the two parts of a 6-pulse gate. This results in a change of the gate phase (see Appendix~\ref{App:arbphase}), which can also be seen as a conditional infidelity ($\mc{C}\neq 1$). The protocol in that case is not robust to the phase shift.

For Protocol \textrm{II}.a and Protocol III, where this finite phase shift alone does not cause errors, one can also consider how much the conditional and leakage-robustness against other errors are affected by it. In fact, this phase fluctuation does not compromise the conditional robustness, but this phenomenon affects the leakage robustness since the phase jump which ensures destructive interference of the leakage amplitudes is effectively modified. However, this effect is a higher-order one: the population left in the $\ket{r}$ state between the two gates is quadratic, so the small correction added by the phase shift gives a high-order correction to the fidelity.

From this we see that, when the positional phase fluctuation is not dominated by the other errors, Protocols \textrm{II}.a and \textrm{III}, which are robust to this error, are more advantageous than Protocols \textrm{I}.a and \textrm{II}.b, which are not. On the other hand, if this effect is suppressed, Protocols \textrm{I}.a and \textrm{II}.b would have the advantage of their shorter execution time.

\section{Expressions of the power series for \texorpdfstring{$\mc{F}$}{F}, \texorpdfstring{$\mc{P}$}{P} and \texorpdfstring{$\mc{C}$}{C}} \label{App:tableanalytic}
In the following, we present the analytical expressions of the expansions shown in Table \ref{tab:tabfids}, as well as the fidelity expansions of the protocols introduced in the main text and appendix, for symmetric and antisymmetric Doppler-induced frequency errors.

To make the analytical expressions lighter, we introduced the notations (denoting $C_n\equiv {\rm cos}\lr{\frac{n\pi}{\sqrt{2}}}$):
\begin{align*}
    A&\equiv C_1^2+C_1+1\\
    B&\equiv 13+4C_1+8C_2-4C_3+3C_4\\
    D&\equiv 30+30C_1+27C_2+2C_3+6C_4+C_6.
\end{align*}
The values presented in Tables \ref{tab:expansionssymdopp} and \ref{tab:expansionsantisymdopp} can similarly be expressed analytically. We present numerical values here to ease comparison between different protocols.

\onecolumngrid

\begin{table*}[ht]
\centering
\begin{tabular}{ |c|c|c|c|c| }
 \hline
 CZ gate protocol & $\mc{F}$ & $\mc{P}$ & $\mc{C}$ & Execution time \\ 
 \hline
 Jaksch et al. \cite{jakschFastQuantumGates2000b} & $1-\frac{\pi^2}{2}\epsilon^2+O\lr{\epsilon^4}$ & $1-\frac{\pi^2}{2}\epsilon^2+O\lr{\epsilon^4}$ & {\color{Green}$1-\frac{\pi^4}{20}\epsilon^4+O\lr{\epsilon^6}$} & $T_{J}\equiv4\pi/\Omega=12.57/\Omega$\\
 \hline
 Protocol \textrm{I} & $1-\frac{A\pi^2}{4} \epsilon^2+O\lr{\epsilon^3}$ & $1-\frac{A\pi^2}{4} \epsilon^2+O\lr{\epsilon^3}$ & {\color{Green}$1-\frac{\pi^4}{640} B\epsilon^4+O\lr{\epsilon^5}$} & $T_{\textrm{I}}\equiv(2+\sqrt{2})\pi/\Omega=10.73/\Omega$\\
 \hline
 Protocol \textrm{II} & {\color{Green}$1-\frac{\pi^4}{640} B\epsilon^4+O\lr{\epsilon^5}$} & {\color{Green}$1-\frac{\pi^6}{1024} D\epsilon^6+O\lr{\epsilon^7}$} & {\color{Green}$1-\frac{\pi^4}{640} B\epsilon^4+O\lr{\epsilon^5}$} & $T_{\textrm{II}}\equiv2(2+\sqrt{2})\pi/\Omega=21.45/\Omega$\\
 \hline
\end{tabular}
\caption{Analytical expressions of the fidelity expansions given in Table \ref{tab:tabfids} in the main text.}
\end{table*}


\begin{table*}[ht]\label{tab:expansionssymdopp}
\centering
\begin{tabular}{ |c|c|c|c|c| }
 \hline
 CZ gate protocol & $\mc{F}$ & $\mc{P}$ & $\mc{C}$ & Execution time \\ 
 \hline
 Jaksch et al. \cite{jakschFastQuantumGates2000b} & $1-2.480\epsilon^2+O\lr{\epsilon^4}$ & $1-\epsilon^2+O\lr{\epsilon^4}$ & $1-1.480\epsilon^2+O\lr{\epsilon^6}$ & $12.57/\Omega$ \\ 
 Levine et al. \cite{levineParallelImplementationHighFidelity2019} & $1-3.000\epsilon^2+O\lr{\epsilon^3}$ & $1-0.077\epsilon^2+O\lr{\epsilon^3}$ & $1-2.923\epsilon^2+O\lr{\epsilon^3}$ & $8.59/\Omega$\\ 
 \hline
 Protocol \textrm{I} & $1-4.314\epsilon^2+O\lr{\epsilon^4}$ & {\color{Green}$1-3.124\epsilon^4+O\lr{\epsilon^6}$} & $1-4.314\epsilon^2+O\lr{\epsilon^4}$ & $10.73/\Omega$\\
 Protocol \textrm{II} &  $1-17.256\epsilon^2+O\lr{\epsilon^4}$ & {\color{Green}$1-6.249\epsilon^4+O\lr{\epsilon^6}$} &  $1-17.256\epsilon^2+O\lr{\epsilon^4}$ & $21.45/\Omega$ \\
 Protocol \textrm{I}.a \eqref{eq:variant1} & $1-2.018 \epsilon^2+O\lr{\epsilon^4}$ & $1-2.018 \epsilon^2+O\lr{\epsilon^4}$ & {\color{Green}$1-0.786\epsilon^4+O\lr{\epsilon^6}$} & $10.73/\Omega$ \\
 Protocol \textrm{I}.a \eqref{eq:variant2} & $1-4.091 \epsilon^2+O\lr{\epsilon^4}$ & $1-4.091 \epsilon^2+O\lr{\epsilon^4}$ & {\color{Green}$1-2.011\epsilon^4+O\lr{\epsilon^6}$} & $10.73/\Omega$ \\
 Protocol \textrm{II}.a \eqref{eq:variant1} & {\color{Green} $1-7.035\epsilon^4+O\lr{\epsilon^6}$} & {\color{Green}$1-6.249\epsilon^4+O\lr{\epsilon^6}$} & {\color{Green} $1-0.786\epsilon^4+O\lr{\epsilon^6}$} & $21.45/\Omega$ \\
 Protocol \textrm{II}.a \eqref{eq:variant2} & {\color{Green} $1-8.260\epsilon^4+O\lr{\epsilon^6}$} & {\color{Green}$1-6.249\epsilon^4+O\lr{\epsilon^6}$} & {\color{Green} $1-2.011\epsilon^4+O\lr{\epsilon^6}$} & $21.45/\Omega$ \\
 Protocol \textrm{II}.b \eqref{eq:variant1} & {\color{Green}$1-0.786\epsilon^4+O\lr{\epsilon^6}$} & {\color{Green}$1-O\lr{\epsilon^6}$} & {\color{Green}$1-0.786\epsilon^4+O\lr{\epsilon^6}$} & $21.45/\Omega$ \\
 Protocol \textrm{II}.b \eqref{eq:variant2} & {\color{Green}$1-2.011\epsilon^4+O\lr{\epsilon^6}$} & {\color{Green}$1-O\lr{\epsilon^6}$} & {\color{Green}$1-2.011\epsilon^4+O\lr{\epsilon^6}$} & $21.45/\Omega$ \\
 Protocol \textrm{III} \eqref{eq:variant1} & {\color{Green}$1-1.570\epsilon^4+O\lr{\epsilon^6}$} & {\color{Green}$1-O\lr{\epsilon^6}$} & {\color{Green}$1-1.570\epsilon^4+O\lr{\epsilon^6}$} & $42.90/\Omega$ \\
 Protocol \textrm{III} \eqref{eq:variant2} & {\color{Green}$1-4.021\epsilon^4+O\lr{\epsilon^6}$} & {\color{Green}$1-O\lr{\epsilon^6}$} & {\color{Green}$1-4.021\epsilon^4+O\lr{\epsilon^6}$} & $42.90/\Omega$ \\
 \hline
\end{tabular}
\caption{Leading-order expansions of $\mc{F}$, $\mc{P}$ and $\mc{C}$ for symmetric detuning fluctuation ($\epsilon=\delta\Delta/\Omega$). Expressions are no longer always the same if we choose sequence \eqref{eq:variant1} or \eqref{eq:variant2} as our basis for pulse area values.}
\label{tab:tabfids_det}
\end{table*}


\begin{table*}[ht]\label{tab:expansionsantisymdopp}
\centering
\begin{tabular}{ |c|c|c|c|c| }
 \hline
 CZ gate protocol & $\mc{F}$ & $\mc{P}$ & $\mc{C}$ & Execution time \\ 
 \hline
 Jaksch et al. \cite{jakschFastQuantumGates2000b} & $1-6.428\epsilon^2+O\lr{\epsilon^4}$ & $1-\epsilon^2+O\lr{\epsilon^4}$ & $1-5.428\epsilon^2+O\lr{\epsilon^4}$ & $12.57/\Omega$ \\
 
 Levine et al. \cite{levineParallelImplementationHighFidelity2019} & $1-11.772\epsilon^2+O\lr{\epsilon^4}$ & $1-3.417\epsilon^2+O\lr{\epsilon^4}$ & $1-8.355\epsilon^2+O\lr{\epsilon^4}$ & $8.59/\Omega$ \\
 \hline
 Protocol \textrm{I} \eqref{eq:variant1} &  $1-17.637\epsilon^2+O\lr{\epsilon^4}$ & $1-6.132\epsilon^2+O\lr{\epsilon^4}$ &  $1-11.505\epsilon^2+O\lr{\epsilon^4}$ & $10.73/\Omega$ \\
 
  Protocol \textrm{I} \eqref{eq:variant2} &  $1-19.313\epsilon^2+O\lr{\epsilon^4}$ & $1-7.808\epsilon^2+O\lr{\epsilon^4}$ &  $1-11.505\epsilon^2+O\lr{\epsilon^4}$ & $10.73/\Omega$ \\
 
 Protocol \textrm{II} \eqref{eq:variant1} &  $1-58.284\epsilon^2+O\lr{\epsilon^4}$ & $1-12.264\epsilon^2+O\lr{\epsilon^4}$ &  $1-46.020\epsilon^2+O\lr{\epsilon^4}$ & $21.45/\Omega$ \\
 Protocol \textrm{II} \eqref{eq:variant2} &  $1-61.636\epsilon^2+O\lr{\epsilon^4}$ & $1-15.616\epsilon^2+O\lr{\epsilon^4}$ &  $1-46.020\epsilon^2+O\lr{\epsilon^4}$ & $21.45/\Omega$ \\
 Protocol \textrm{I}.a \eqref{eq:variant1} &  $1-2\epsilon^2+O\lr{\epsilon^4}$ & $1-2\epsilon^2+O\lr{\epsilon^4}$ & {\color{Green} $1-\frac{4}{5}\epsilon^4+O\lr{\epsilon^6}$} & $10.73/\Omega$ \\
 
 Protocol \textrm{I}.a \eqref{eq:variant2} &  $1-3.591\epsilon^2+O\lr{\epsilon^4}$ & $1-3.591\epsilon^2+O\lr{\epsilon^4}$ & {\color{Green} $1-2.580\epsilon^4+O\lr{\epsilon^6}$} & $10.73/\Omega$ \\
 
 Protocol \textrm{II}.a \eqref{eq:variant1} &  $1-12.264\epsilon^2+O\lr{\epsilon^4}$ &  $1-12.264\epsilon^2+O\lr{\epsilon^4}$ & {\color{Green} $1-171.462\epsilon^4+O\lr{\epsilon^6}$} & $21.45/\Omega$ \\
 
 Protocol \textrm{II}.a \eqref{eq:variant2} &  $1-15.616\epsilon^2+O\lr{\epsilon^4}$ &  $1-15.616\epsilon^2+O\lr{\epsilon^4}$ & {\color{Green} $1-272.779\epsilon^4+O\lr{\epsilon^6}$} & $21.45/\Omega$ \\
 
 Protocol \textrm{II}.b \eqref{eq:variant1} & {\color{Green} $1-\frac{4}{5}\epsilon^4+O\lr{\epsilon^6}$} & {\color{Green}$ 1-O\lr{\epsilon^6}$} & {\color{Green} $1-\frac{4}{5}\epsilon^4+O\lr{\epsilon^6}$} & $21.45/\Omega$ \\
 
 Protocol \textrm{II}.b \eqref{eq:variant2} & {\color{Green} $1-2.580\epsilon^4+O\lr{\epsilon^6}$} & {\color{Green}$1-O\lr{\epsilon^6}$} & {\color{Green} $1-2.580\epsilon^4+O\lr{\epsilon^6}$} & $21.45/\Omega$ \\
 
 Protocol \textrm{III} \eqref{eq:variant1} & {\color{Green}$1-676\epsilon^4+O\lr{\epsilon^6}$} & {\color{Green}$1-513\epsilon^4+O\lr{\epsilon^6}$} & {\color{Green}$1-163\epsilon^4+O\lr{\epsilon^6}$} & $42.90/\Omega$ \\
 
 Protocol \textrm{III} \eqref{eq:variant2} & {\color{Green}$1-1090\epsilon^4+O\lr{\epsilon^6}$} & {\color{Green}$1-837\epsilon^4+O\lr{\epsilon^6}$} & {\color{Green}$1-253\epsilon^4+O\lr{\epsilon^6}$} & $42.90/\Omega$ \\
 \hline
\end{tabular}
\caption{Leading-order expansions of $\mc{F}$, $\mc{P}$ and $\mc{C}$ for antisymmetric detuning fluctuations ($\epsilon=\frac{\delta\Delta_1}{\Omega}=-\frac{\delta\Delta_2}{\Omega}$).}
\label{tab:tabfids_inhomo}
\end{table*}

\clearpage
\twocolumngrid

\bibliographystyle{apsrev4-2-mod.bst}
\bibliography{biblio.bib,HannesBib.bib}

\begin{thebibliography}{34}%
\makeatletter
\providecommand \@ifxundefined [1]{%
 \@ifx{#1\undefined}
}%
\providecommand \@ifnum [1]{%
 \ifnum #1\expandafter \@firstoftwo
 \else \expandafter \@secondoftwo
 \fi
}%
\providecommand \@ifx [1]{%
 \ifx #1\expandafter \@firstoftwo
 \else \expandafter \@secondoftwo
 \fi
}%
\providecommand \natexlab [1]{#1}%
\providecommand \enquote  [1]{``#1''}%
\providecommand \bibnamefont  [1]{#1}%
\providecommand \bibfnamefont [1]{#1}%
\providecommand \citenamefont [1]{#1}%
\providecommand \href@noop [0]{\@secondoftwo}%
\providecommand \href [0]{\begingroup \@sanitize@url \@href}%
\providecommand \@href[1]{\@@startlink{#1}\@@href}%
\providecommand \@@href[1]{\endgroup#1\@@endlink}%
\providecommand \@sanitize@url [0]{\catcode `\\12\catcode `\$12\catcode
  `\&12\catcode `\#12\catcode `\^12\catcode `\_12\catcode `\%12\relax}%
\providecommand \@@startlink[1]{}%
\providecommand \@@endlink[0]{}%
\providecommand \url  [0]{\begingroup\@sanitize@url \@url }%
\providecommand \@url [1]{\endgroup\@href {#1}{\urlprefix }}%
\providecommand \urlprefix  [0]{URL }%
\providecommand \Eprint [0]{\href }%
\providecommand \doibase [0]{https://doi.org/}%
\providecommand \selectlanguage [0]{\@gobble}%
\providecommand \bibinfo  [0]{\@secondoftwo}%
\providecommand \bibfield  [0]{\@secondoftwo}%
\providecommand \translation [1]{[#1]}%
\providecommand \BibitemOpen [0]{}%
\providecommand \bibitemStop [0]{}%
\providecommand \bibitemNoStop [0]{.\EOS\space}%
\providecommand \EOS [0]{\spacefactor3000\relax}%
\providecommand \BibitemShut  [1]{\csname bibitem#1\endcsname}%
\let\auto@bib@innerbib\@empty
\bibitem [{\citenamefont {Browaeys}\ and\ \citenamefont
  {Lahaye}(2020)}]{browaeysManybodyPhysicsIndividually2020}%
  \BibitemOpen
  \bibfield  {author} {\bibinfo {author} {\bibfnamefont {A.}~\bibnamefont
  {Browaeys}}\ and\ \bibinfo {author} {\bibfnamefont {T.}~\bibnamefont
  {Lahaye}},\ }\bibfield  {title} {\emph {\bibinfo {title} {Many-body physics
  with individually controlled {{Rydberg}} atoms}},\ }\href
  {https://doi.org/10.1038/s41567-019-0733-z} {\bibfield  {journal} {\bibinfo
  {journal} {Nature Physics}\ }\textbf {\bibinfo {volume} {16}},\ \bibinfo
  {pages} {132} (\bibinfo {year} {2020})}\BibitemShut {NoStop}%
\bibitem [{\citenamefont {Weimer}\ \emph {et~al.}(2010)\citenamefont {Weimer},
  \citenamefont {M{\"u}ller}, \citenamefont {Lesanovsky}, \citenamefont
  {Zoller},\ and\ \citenamefont
  {B{\"u}chler}}]{weimerRydbergQuantumSimulator2010a}%
  \BibitemOpen
  \bibfield  {author} {\bibinfo {author} {\bibfnamefont {H.}~\bibnamefont
  {Weimer}}, \bibinfo {author} {\bibfnamefont {M.}~\bibnamefont {M{\"u}ller}},
  \bibinfo {author} {\bibfnamefont {I.}~\bibnamefont {Lesanovsky}}, \bibinfo
  {author} {\bibfnamefont {P.}~\bibnamefont {Zoller}},\ and\ \bibinfo {author}
  {\bibfnamefont {H.~P.}\ \bibnamefont {B{\"u}chler}},\ }\bibfield  {title}
  {\emph {\bibinfo {title} {A {{Rydberg}} quantum simulator}},\ }\href
  {https://doi.org/10.1038/nphys1614} {\bibfield  {journal} {\bibinfo
  {journal} {Nature Physics}\ }\textbf {\bibinfo {volume} {6}},\ \bibinfo
  {pages} {382} (\bibinfo {year} {2010})}\BibitemShut {NoStop}%
\bibitem [{\citenamefont {Bernien}\ \emph {et~al.}(2017)\citenamefont
  {Bernien}, \citenamefont {Schwartz}, \citenamefont {Keesling}, \citenamefont
  {Levine}, \citenamefont {Omran}, \citenamefont {Pichler}, \citenamefont
  {Choi}, \citenamefont {Zibrov}, \citenamefont {Endres}, \citenamefont
  {Greiner}, \citenamefont {Vuleti{\'c}},\ and\ \citenamefont
  {Lukin}}]{bernienProbingManybodyDynamics2017a}%
  \BibitemOpen
  \bibfield  {author} {\bibinfo {author} {\bibfnamefont {H.}~\bibnamefont
  {Bernien}}, \bibinfo {author} {\bibfnamefont {S.}~\bibnamefont {Schwartz}},
  \bibinfo {author} {\bibfnamefont {A.}~\bibnamefont {Keesling}}, \bibinfo
  {author} {\bibfnamefont {H.}~\bibnamefont {Levine}}, \bibinfo {author}
  {\bibfnamefont {A.}~\bibnamefont {Omran}}, \bibinfo {author} {\bibfnamefont
  {H.}~\bibnamefont {Pichler}}, \bibinfo {author} {\bibfnamefont
  {S.}~\bibnamefont {Choi}}, \bibinfo {author} {\bibfnamefont {A.~S.}\
  \bibnamefont {Zibrov}}, \bibinfo {author} {\bibfnamefont {M.}~\bibnamefont
  {Endres}}, \bibinfo {author} {\bibfnamefont {M.}~\bibnamefont {Greiner}},
  \bibinfo {author} {\bibfnamefont {V.}~\bibnamefont {Vuleti{\'c}}},\ and\
  \bibinfo {author} {\bibfnamefont {M.~D.}\ \bibnamefont {Lukin}},\ }\bibfield
  {title} {\emph {\bibinfo {title} {Probing many-body dynamics on a 51-atom
  quantum simulator}},\ }\href {https://doi.org/10.1038/nature24622} {\bibfield
   {journal} {\bibinfo  {journal} {Nature}\ }\textbf {\bibinfo {volume}
  {551}},\ \bibinfo {pages} {579} (\bibinfo {year} {2017})}\BibitemShut
  {NoStop}%
\bibitem [{\citenamefont {Labuhn}\ \emph {et~al.}(2016)\citenamefont {Labuhn},
  \citenamefont {Barredo}, \citenamefont {Ravets}, \citenamefont {{de
  L{\'e}s{\'e}leuc}}, \citenamefont {Macr{\`i}}, \citenamefont {Lahaye},\ and\
  \citenamefont {Browaeys}}]{labuhnTunableTwodimensionalArrays2016}%
  \BibitemOpen
  \bibfield  {author} {\bibinfo {author} {\bibfnamefont {H.}~\bibnamefont
  {Labuhn}}, \bibinfo {author} {\bibfnamefont {D.}~\bibnamefont {Barredo}},
  \bibinfo {author} {\bibfnamefont {S.}~\bibnamefont {Ravets}}, \bibinfo
  {author} {\bibfnamefont {S.}~\bibnamefont {{de L{\'e}s{\'e}leuc}}}, \bibinfo
  {author} {\bibfnamefont {T.}~\bibnamefont {Macr{\`i}}}, \bibinfo {author}
  {\bibfnamefont {T.}~\bibnamefont {Lahaye}},\ and\ \bibinfo {author}
  {\bibfnamefont {A.}~\bibnamefont {Browaeys}},\ }\bibfield  {title} {\emph
  {\bibinfo {title} {Tunable two-dimensional arrays of single {{Rydberg}} atoms
  for realizing quantum {{Ising}} models}},\ }\href
  {https://doi.org/10.1038/nature18274} {\bibfield  {journal} {\bibinfo
  {journal} {Nature}\ }\textbf {\bibinfo {volume} {534}},\ \bibinfo {pages}
  {667} (\bibinfo {year} {2016})}\BibitemShut {NoStop}%
\bibitem [{\citenamefont {Jaksch}\ \emph {et~al.}(2000)\citenamefont {Jaksch},
  \citenamefont {Cirac}, \citenamefont {Zoller}, \citenamefont {Rolston},
  \citenamefont {C{\^o}t{\'e}},\ and\ \citenamefont
  {Lukin}}]{jakschFastQuantumGates2000b}%
  \BibitemOpen
  \bibfield  {author} {\bibinfo {author} {\bibfnamefont {D.}~\bibnamefont
  {Jaksch}}, \bibinfo {author} {\bibfnamefont {J.~I.}\ \bibnamefont {Cirac}},
  \bibinfo {author} {\bibfnamefont {P.}~\bibnamefont {Zoller}}, \bibinfo
  {author} {\bibfnamefont {S.~L.}\ \bibnamefont {Rolston}}, \bibinfo {author}
  {\bibfnamefont {R.}~\bibnamefont {C{\^o}t{\'e}}},\ and\ \bibinfo {author}
  {\bibfnamefont {M.~D.}\ \bibnamefont {Lukin}},\ }\bibfield  {title} {\emph
  {\bibinfo {title} {Fast {{Quantum Gates}} for {{Neutral Atoms}}}},\ }\href
  {https://doi.org/10.1103/PhysRevLett.85.2208} {\bibfield  {journal} {\bibinfo
   {journal} {Physical Review Letters}\ }\textbf {\bibinfo {volume} {85}},\
  \bibinfo {pages} {2208} (\bibinfo {year} {2000})}\BibitemShut {NoStop}%
\bibitem [{\citenamefont {Lukin}\ \emph {et~al.}(2001)\citenamefont {Lukin},
  \citenamefont {Fleischhauer}, \citenamefont {Cote}, \citenamefont {Duan},
  \citenamefont {Jaksch}, \citenamefont {Cirac},\ and\ \citenamefont
  {Zoller}}]{lukinDipoleBlockadeQuantum2001}%
  \BibitemOpen
  \bibfield  {author} {\bibinfo {author} {\bibfnamefont {M.~D.}\ \bibnamefont
  {Lukin}}, \bibinfo {author} {\bibfnamefont {M.}~\bibnamefont {Fleischhauer}},
  \bibinfo {author} {\bibfnamefont {R.}~\bibnamefont {Cote}}, \bibinfo {author}
  {\bibfnamefont {L.~M.}\ \bibnamefont {Duan}}, \bibinfo {author}
  {\bibfnamefont {D.}~\bibnamefont {Jaksch}}, \bibinfo {author} {\bibfnamefont
  {J.~I.}\ \bibnamefont {Cirac}},\ and\ \bibinfo {author} {\bibfnamefont
  {P.}~\bibnamefont {Zoller}},\ }\bibfield  {title} {\emph {\bibinfo {title}
  {Dipole {{Blockade}} and {{Quantum Information Processing}} in {{Mesoscopic
  Atomic Ensembles}}}},\ }\href {https://doi.org/10.1103/PhysRevLett.87.037901}
  {\bibfield  {journal} {\bibinfo  {journal} {Physical Review Letters}\
  }\textbf {\bibinfo {volume} {87}},\ \bibinfo {pages} {037901} (\bibinfo
  {year} {2001})}\BibitemShut {NoStop}%
\bibitem [{\citenamefont {Urban}\ \emph {et~al.}(2009)\citenamefont {Urban},
  \citenamefont {Johnson}, \citenamefont {Henage}, \citenamefont {Isenhower},
  \citenamefont {Yavuz}, \citenamefont {Walker},\ and\ \citenamefont
  {Saffman}}]{urbanObservationrydberg}%
  \BibitemOpen
  \bibfield  {author} {\bibinfo {author} {\bibfnamefont {E.}~\bibnamefont
  {Urban}}, \bibinfo {author} {\bibfnamefont {T.}~\bibnamefont {Johnson}},
  \bibinfo {author} {\bibfnamefont {T.}~\bibnamefont {Henage}}, \bibinfo
  {author} {\bibfnamefont {L.}~\bibnamefont {Isenhower}}, \bibinfo {author}
  {\bibfnamefont {D.}~\bibnamefont {Yavuz}}, \bibinfo {author} {\bibfnamefont
  {T.}~\bibnamefont {Walker}},\ and\ \bibinfo {author} {\bibfnamefont
  {M.}~\bibnamefont {Saffman}},\ }\bibfield  {title} {\emph {\bibinfo {title}
  {Observation of {{Rydberg}} blockade between two atoms}},\ }\href
  {https://doi.org/10.1038/nphys1178} {\bibfield  {journal} {\bibinfo
  {journal} {Nature Physics}\ }\textbf {\bibinfo {volume} {5}},\ \bibinfo
  {pages} {110} (\bibinfo {year} {2009})}\BibitemShut {NoStop}%
\bibitem [{\citenamefont {Wilk}\ \emph {et~al.}(2010)\citenamefont {Wilk},
  \citenamefont {Ga\"etan}, \citenamefont {Evellin}, \citenamefont {Wolters},
  \citenamefont {Miroshnychenko}, \citenamefont {Grangier},\ and\ \citenamefont
  {Browaeys}}]{wilkEntanglementTwoIndividual2010_2}%
  \BibitemOpen
  \bibfield  {author} {\bibinfo {author} {\bibfnamefont {T.}~\bibnamefont
  {Wilk}}, \bibinfo {author} {\bibfnamefont {A.}~\bibnamefont {Ga\"etan}},
  \bibinfo {author} {\bibfnamefont {C.}~\bibnamefont {Evellin}}, \bibinfo
  {author} {\bibfnamefont {J.}~\bibnamefont {Wolters}}, \bibinfo {author}
  {\bibfnamefont {Y.}~\bibnamefont {Miroshnychenko}}, \bibinfo {author}
  {\bibfnamefont {P.}~\bibnamefont {Grangier}},\ and\ \bibinfo {author}
  {\bibfnamefont {A.}~\bibnamefont {Browaeys}},\ }\bibfield  {title} {\emph
  {\bibinfo {title} {Entanglement of {{Two Individual Neutral Atoms Using
  Rydberg Blockade}}}},\ }\href
  {https://doi.org/10.1103/PhysRevLett.104.010502} {\bibfield  {journal}
  {\bibinfo  {journal} {Physical Review Letters}\ }\textbf {\bibinfo {volume}
  {104}},\ \bibinfo {pages} {010502} (\bibinfo {year} {2010})}\BibitemShut
  {NoStop}%
\bibitem [{\citenamefont {Levine}\ \emph {et~al.}(2019)\citenamefont {Levine},
  \citenamefont {Keesling}, \citenamefont {Semeghini}, \citenamefont {Omran},
  \citenamefont {Wang}, \citenamefont {Ebadi}, \citenamefont {Bernien},
  \citenamefont {Greiner}, \citenamefont {Vuleti{\'c}}, \citenamefont
  {Pichler},\ and\ \citenamefont
  {Lukin}}]{levineParallelImplementationHighFidelity2019}%
  \BibitemOpen
  \bibfield  {author} {\bibinfo {author} {\bibfnamefont {H.}~\bibnamefont
  {Levine}}, \bibinfo {author} {\bibfnamefont {A.}~\bibnamefont {Keesling}},
  \bibinfo {author} {\bibfnamefont {G.}~\bibnamefont {Semeghini}}, \bibinfo
  {author} {\bibfnamefont {A.}~\bibnamefont {Omran}}, \bibinfo {author}
  {\bibfnamefont {T.~T.}\ \bibnamefont {Wang}}, \bibinfo {author}
  {\bibfnamefont {S.}~\bibnamefont {Ebadi}}, \bibinfo {author} {\bibfnamefont
  {H.}~\bibnamefont {Bernien}}, \bibinfo {author} {\bibfnamefont
  {M.}~\bibnamefont {Greiner}}, \bibinfo {author} {\bibfnamefont
  {V.}~\bibnamefont {Vuleti{\'c}}}, \bibinfo {author} {\bibfnamefont
  {H.}~\bibnamefont {Pichler}},\ and\ \bibinfo {author} {\bibfnamefont {M.~D.}\
  \bibnamefont {Lukin}},\ }\bibfield  {title} {\emph {\bibinfo {title}
  {Parallel {{Implementation}} of {{High-Fidelity Multiqubit Gates}} with
  {{Neutral Atoms}}}},\ }\href {https://doi.org/10.1103/PhysRevLett.123.170503}
  {\bibfield  {journal} {\bibinfo  {journal} {Physical Review Letters}\
  }\textbf {\bibinfo {volume} {123}},\ \bibinfo {pages} {170503} (\bibinfo
  {year} {2019})}\BibitemShut {NoStop}%
\bibitem [{\citenamefont {Kaufman}\ and\ \citenamefont
  {Ni}(2021)}]{kaufmanQuantumScienceOptical2021}%
  \BibitemOpen
  \bibfield  {author} {\bibinfo {author} {\bibfnamefont {A.~M.}\ \bibnamefont
  {Kaufman}}\ and\ \bibinfo {author} {\bibfnamefont {K.-K.}\ \bibnamefont
  {Ni}},\ }\bibfield  {title} {\emph {\bibinfo {title} {Quantum science with
  optical tweezer arrays of ultracold atoms and molecules}},\ }\href
  {https://doi.org/10.1038/s41567-021-01357-2} {\bibfield  {journal} {\bibinfo
  {journal} {Nature Physics}\ }\textbf {\bibinfo {volume} {17}},\ \bibinfo
  {pages} {1324} (\bibinfo {year} {2021})}\BibitemShut {NoStop}%
\bibitem [{\citenamefont {Graham}\ \emph {et~al.}(2022)\citenamefont {Graham},
  \citenamefont {Song}, \citenamefont {Scott}, \citenamefont {Poole},
  \citenamefont {Phuttitarn}, \citenamefont {Jooya}, \citenamefont {Eichler},
  \citenamefont {Jiang}, \citenamefont {Marra}, \citenamefont {Grinkemeyer},
  \citenamefont {Kwon}, \citenamefont {Ebert}, \citenamefont {Cherek},
  \citenamefont {Lichtman}, \citenamefont {Gillette}, \citenamefont {Gilbert},
  \citenamefont {Bowman}, \citenamefont {Ballance}, \citenamefont {Campbell},
  \citenamefont {Dahl}, \citenamefont {Crawford}, \citenamefont {Blunt},
  \citenamefont {Rogers}, \citenamefont {Noel},\ and\ \citenamefont
  {Saffman}}]{grahamMultiqubitEntanglementAlgorithms2022}%
  \BibitemOpen
  \bibfield  {author} {\bibinfo {author} {\bibfnamefont {T.~M.}\ \bibnamefont
  {Graham}}, \bibinfo {author} {\bibfnamefont {Y.}~\bibnamefont {Song}},
  \bibinfo {author} {\bibfnamefont {J.}~\bibnamefont {Scott}}, \bibinfo
  {author} {\bibfnamefont {C.}~\bibnamefont {Poole}}, \bibinfo {author}
  {\bibfnamefont {L.}~\bibnamefont {Phuttitarn}}, \bibinfo {author}
  {\bibfnamefont {K.}~\bibnamefont {Jooya}}, \bibinfo {author} {\bibfnamefont
  {P.}~\bibnamefont {Eichler}}, \bibinfo {author} {\bibfnamefont
  {X.}~\bibnamefont {Jiang}}, \bibinfo {author} {\bibfnamefont
  {A.}~\bibnamefont {Marra}}, \bibinfo {author} {\bibfnamefont
  {B.}~\bibnamefont {Grinkemeyer}}, \bibinfo {author} {\bibfnamefont
  {M.}~\bibnamefont {Kwon}}, \bibinfo {author} {\bibfnamefont {M.}~\bibnamefont
  {Ebert}}, \bibinfo {author} {\bibfnamefont {J.}~\bibnamefont {Cherek}},
  \bibinfo {author} {\bibfnamefont {M.~T.}\ \bibnamefont {Lichtman}}, \bibinfo
  {author} {\bibfnamefont {M.}~\bibnamefont {Gillette}}, \bibinfo {author}
  {\bibfnamefont {J.}~\bibnamefont {Gilbert}}, \bibinfo {author} {\bibfnamefont
  {D.}~\bibnamefont {Bowman}}, \bibinfo {author} {\bibfnamefont
  {T.}~\bibnamefont {Ballance}}, \bibinfo {author} {\bibfnamefont
  {C.}~\bibnamefont {Campbell}}, \bibinfo {author} {\bibfnamefont {E.~D.}\
  \bibnamefont {Dahl}}, \bibinfo {author} {\bibfnamefont {O.}~\bibnamefont
  {Crawford}}, \bibinfo {author} {\bibfnamefont {N.~S.}\ \bibnamefont {Blunt}},
  \bibinfo {author} {\bibfnamefont {B.}~\bibnamefont {Rogers}}, \bibinfo
  {author} {\bibfnamefont {T.}~\bibnamefont {Noel}},\ and\ \bibinfo {author}
  {\bibfnamefont {M.}~\bibnamefont {Saffman}},\ }\bibfield  {title} {\emph
  {\bibinfo {title} {Multi-qubit entanglement and algorithms on a neutral-atom
  quantum computer}},\ }\href {https://doi.org/10.1038/s41586-022-04603-6}
  {\bibfield  {journal} {\bibinfo  {journal} {Nature}\ }\textbf {\bibinfo
  {volume} {604}},\ \bibinfo {pages} {457} (\bibinfo {year}
  {2022})}\BibitemShut {NoStop}%
\bibitem [{\citenamefont {Ma}\ \emph {et~al.}(2022)\citenamefont {Ma},
  \citenamefont {Burgers}, \citenamefont {Liu}, \citenamefont {Wilson},
  \citenamefont {Zhang},\ and\ \citenamefont
  {Thompson}}]{maUniversalGateOperations2022}%
  \BibitemOpen
  \bibfield  {author} {\bibinfo {author} {\bibfnamefont {S.}~\bibnamefont
  {Ma}}, \bibinfo {author} {\bibfnamefont {A.~P.}\ \bibnamefont {Burgers}},
  \bibinfo {author} {\bibfnamefont {G.}~\bibnamefont {Liu}}, \bibinfo {author}
  {\bibfnamefont {J.}~\bibnamefont {Wilson}}, \bibinfo {author} {\bibfnamefont
  {B.}~\bibnamefont {Zhang}},\ and\ \bibinfo {author} {\bibfnamefont {J.~D.}\
  \bibnamefont {Thompson}},\ }\bibfield  {title} {\emph {\bibinfo {title}
  {Universal {{Gate Operations}} on {{Nuclear Spin Qubits}} in an {{Optical
  Tweezer Array}} of {{171Yb Atoms}}}},\ }\href
  {https://doi.org/10.1103/PhysRevX.12.021028} {\bibfield  {journal} {\bibinfo
  {journal} {Physical Review X}\ }\textbf {\bibinfo {volume} {12}},\ \bibinfo
  {pages} {021028} (\bibinfo {year} {2022})}\BibitemShut {NoStop}%
\bibitem [{\citenamefont {Madjarov}\ \emph {et~al.}(2020)\citenamefont
  {Madjarov}, \citenamefont {Covey}, \citenamefont {Shaw}, \citenamefont
  {Choi}, \citenamefont {Kale}, \citenamefont {Cooper}, \citenamefont
  {Pichler}, \citenamefont {Schkolnik}, \citenamefont {Williams},\ and\
  \citenamefont {Endres}}]{madjarovHighfidelityEntanglementDetection2020a}%
  \BibitemOpen
  \bibfield  {author} {\bibinfo {author} {\bibfnamefont {I.~S.}\ \bibnamefont
  {Madjarov}}, \bibinfo {author} {\bibfnamefont {J.~P.}\ \bibnamefont {Covey}},
  \bibinfo {author} {\bibfnamefont {A.~L.}\ \bibnamefont {Shaw}}, \bibinfo
  {author} {\bibfnamefont {J.}~\bibnamefont {Choi}}, \bibinfo {author}
  {\bibfnamefont {A.}~\bibnamefont {Kale}}, \bibinfo {author} {\bibfnamefont
  {A.}~\bibnamefont {Cooper}}, \bibinfo {author} {\bibfnamefont
  {H.}~\bibnamefont {Pichler}}, \bibinfo {author} {\bibfnamefont
  {V.}~\bibnamefont {Schkolnik}}, \bibinfo {author} {\bibfnamefont {J.~R.}\
  \bibnamefont {Williams}},\ and\ \bibinfo {author} {\bibfnamefont
  {M.}~\bibnamefont {Endres}},\ }\bibfield  {title} {\emph {\bibinfo {title}
  {High-fidelity entanglement and detection of alkaline-earth {{Rydberg}}
  atoms}},\ }\href {https://doi.org/10.1038/s41567-020-0903-z} {\bibfield
  {journal} {\bibinfo  {journal} {Nature Physics}\ }\textbf {\bibinfo {volume}
  {16}},\ \bibinfo {pages} {857} (\bibinfo {year} {2020})}\BibitemShut
  {NoStop}%
\bibitem [{\citenamefont {Bluvstein}\ \emph {et~al.}(2022)\citenamefont
  {Bluvstein}, \citenamefont {Levine}, \citenamefont {Semeghini}, \citenamefont
  {Wang}, \citenamefont {Ebadi}, \citenamefont {Kalinowski}, \citenamefont
  {Keesling}, \citenamefont {Maskara}, \citenamefont {Pichler},\ and\
  \citenamefont {Greiner}}]{bluvsteinQuantumProcessorBased2022_2}%
  \BibitemOpen
  \bibfield  {author} {\bibinfo {author} {\bibfnamefont {D.}~\bibnamefont
  {Bluvstein}}, \bibinfo {author} {\bibfnamefont {H.}~\bibnamefont {Levine}},
  \bibinfo {author} {\bibfnamefont {G.}~\bibnamefont {Semeghini}}, \bibinfo
  {author} {\bibfnamefont {T.~T.}\ \bibnamefont {Wang}}, \bibinfo {author}
  {\bibfnamefont {S.}~\bibnamefont {Ebadi}}, \bibinfo {author} {\bibfnamefont
  {M.}~\bibnamefont {Kalinowski}}, \bibinfo {author} {\bibfnamefont
  {A.}~\bibnamefont {Keesling}}, \bibinfo {author} {\bibfnamefont
  {N.}~\bibnamefont {Maskara}}, \bibinfo {author} {\bibfnamefont
  {H.}~\bibnamefont {Pichler}},\ and\ \bibinfo {author} {\bibfnamefont
  {M.}~\bibnamefont {Greiner}},\ }\bibfield  {title} {\emph {\bibinfo {title}
  {A quantum processor based on coherent transport of entangled atom arrays}},\
  }\href {https://doi.org/10.1038/s41586-022-04592-6} {\bibfield  {journal}
  {\bibinfo  {journal} {Nature}\ }\textbf {\bibinfo {volume} {604}},\ \bibinfo
  {pages} {451} (\bibinfo {year} {2022})}\BibitemShut {NoStop}%
\bibitem [{\citenamefont {Omran}\ \emph {et~al.}(2019)\citenamefont {Omran},
  \citenamefont {Levine}, \citenamefont {Keesling}, \citenamefont {Semeghini},
  \citenamefont {Wang}, \citenamefont {Ebadi}, \citenamefont {Bernien},
  \citenamefont {Zibrov}, \citenamefont {Pichler}, \citenamefont {Choi},
  \citenamefont {Cui}, \citenamefont {Rossignolo}, \citenamefont {Rembold},
  \citenamefont {Montangero}, \citenamefont {Calarco}, \citenamefont {Endres},
  \citenamefont {Greiner}, \citenamefont {Vuleti{\'c}},\ and\ \citenamefont
  {Lukin}}]{omranGenerationManipulationSchrodinger2019a}%
  \BibitemOpen
  \bibfield  {author} {\bibinfo {author} {\bibfnamefont {A.}~\bibnamefont
  {Omran}}, \bibinfo {author} {\bibfnamefont {H.}~\bibnamefont {Levine}},
  \bibinfo {author} {\bibfnamefont {A.}~\bibnamefont {Keesling}}, \bibinfo
  {author} {\bibfnamefont {G.}~\bibnamefont {Semeghini}}, \bibinfo {author}
  {\bibfnamefont {T.~T.}\ \bibnamefont {Wang}}, \bibinfo {author}
  {\bibfnamefont {S.}~\bibnamefont {Ebadi}}, \bibinfo {author} {\bibfnamefont
  {H.}~\bibnamefont {Bernien}}, \bibinfo {author} {\bibfnamefont {A.~S.}\
  \bibnamefont {Zibrov}}, \bibinfo {author} {\bibfnamefont {H.}~\bibnamefont
  {Pichler}}, \bibinfo {author} {\bibfnamefont {S.}~\bibnamefont {Choi}},
  \bibinfo {author} {\bibfnamefont {J.}~\bibnamefont {Cui}}, \bibinfo {author}
  {\bibfnamefont {M.}~\bibnamefont {Rossignolo}}, \bibinfo {author}
  {\bibfnamefont {P.}~\bibnamefont {Rembold}}, \bibinfo {author} {\bibfnamefont
  {S.}~\bibnamefont {Montangero}}, \bibinfo {author} {\bibfnamefont
  {T.}~\bibnamefont {Calarco}}, \bibinfo {author} {\bibfnamefont
  {M.}~\bibnamefont {Endres}}, \bibinfo {author} {\bibfnamefont
  {M.}~\bibnamefont {Greiner}}, \bibinfo {author} {\bibfnamefont
  {V.}~\bibnamefont {Vuleti{\'c}}},\ and\ \bibinfo {author} {\bibfnamefont
  {M.~D.}\ \bibnamefont {Lukin}},\ }\bibfield  {title} {\emph {\bibinfo {title}
  {Generation and manipulation of {{Schr\"odinger}} cat states in {{Rydberg}}
  atom arrays}},\ }\href {https://doi.org/10.1126/science.aax9743} {\bibfield
  {journal} {\bibinfo  {journal} {Science}\ }\textbf {\bibinfo {volume}
  {365}},\ \bibinfo {pages} {570} (\bibinfo {year} {2019})}\BibitemShut
  {NoStop}%
\bibitem [{\citenamefont {Semeghini}\ \emph {et~al.}(2021)\citenamefont
  {Semeghini}, \citenamefont {Levine}, \citenamefont {Keesling}, \citenamefont
  {Ebadi}, \citenamefont {Wang}, \citenamefont {Bluvstein}, \citenamefont
  {Verresen}, \citenamefont {Pichler}, \citenamefont {Kalinowski},
  \citenamefont {Samajdar}, \citenamefont {Omran}, \citenamefont {Sachdev},
  \citenamefont {Vishwanath}, \citenamefont {Greiner}, \citenamefont
  {Vuleti{\'c}},\ and\ \citenamefont
  {Lukin}}]{semeghiniProbingTopologicalSpin2021_2}%
  \BibitemOpen
  \bibfield  {author} {\bibinfo {author} {\bibfnamefont {G.}~\bibnamefont
  {Semeghini}}, \bibinfo {author} {\bibfnamefont {H.}~\bibnamefont {Levine}},
  \bibinfo {author} {\bibfnamefont {A.}~\bibnamefont {Keesling}}, \bibinfo
  {author} {\bibfnamefont {S.}~\bibnamefont {Ebadi}}, \bibinfo {author}
  {\bibfnamefont {T.~T.}\ \bibnamefont {Wang}}, \bibinfo {author}
  {\bibfnamefont {D.}~\bibnamefont {Bluvstein}}, \bibinfo {author}
  {\bibfnamefont {R.}~\bibnamefont {Verresen}}, \bibinfo {author}
  {\bibfnamefont {H.}~\bibnamefont {Pichler}}, \bibinfo {author} {\bibfnamefont
  {M.}~\bibnamefont {Kalinowski}}, \bibinfo {author} {\bibfnamefont
  {R.}~\bibnamefont {Samajdar}}, \bibinfo {author} {\bibfnamefont
  {A.}~\bibnamefont {Omran}}, \bibinfo {author} {\bibfnamefont
  {S.}~\bibnamefont {Sachdev}}, \bibinfo {author} {\bibfnamefont
  {A.}~\bibnamefont {Vishwanath}}, \bibinfo {author} {\bibfnamefont
  {M.}~\bibnamefont {Greiner}}, \bibinfo {author} {\bibfnamefont
  {V.}~\bibnamefont {Vuleti{\'c}}},\ and\ \bibinfo {author} {\bibfnamefont
  {M.~D.}\ \bibnamefont {Lukin}},\ }\bibfield  {title} {\emph {\bibinfo {title}
  {Probing topological spin liquids on a programmable quantum simulator}},\
  }\href {https://doi.org/10.1126/science.abi8794} {\bibfield  {journal}
  {\bibinfo  {journal} {Science}\ }\textbf {\bibinfo {volume} {374}},\ \bibinfo
  {pages} {1242} (\bibinfo {year} {2021})}\BibitemShut {NoStop}%
\bibitem [{\citenamefont {Young}\ \emph {et~al.}(2020)\citenamefont {Young},
  \citenamefont {Eckner}, \citenamefont {Milner}, \citenamefont {Kedar},
  \citenamefont {Norcia}, \citenamefont {Oelker}, \citenamefont {Schine},
  \citenamefont {Ye},\ and\ \citenamefont
  {Kaufman}}]{youngHalfminutescaleAtomicCoherence2020}%
  \BibitemOpen
  \bibfield  {author} {\bibinfo {author} {\bibfnamefont {A.~W.}\ \bibnamefont
  {Young}}, \bibinfo {author} {\bibfnamefont {W.~J.}\ \bibnamefont {Eckner}},
  \bibinfo {author} {\bibfnamefont {W.~R.}\ \bibnamefont {Milner}}, \bibinfo
  {author} {\bibfnamefont {D.}~\bibnamefont {Kedar}}, \bibinfo {author}
  {\bibfnamefont {M.~A.}\ \bibnamefont {Norcia}}, \bibinfo {author}
  {\bibfnamefont {E.}~\bibnamefont {Oelker}}, \bibinfo {author} {\bibfnamefont
  {N.}~\bibnamefont {Schine}}, \bibinfo {author} {\bibfnamefont
  {J.}~\bibnamefont {Ye}},\ and\ \bibinfo {author} {\bibfnamefont {A.~M.}\
  \bibnamefont {Kaufman}},\ }\bibfield  {title} {\emph {\bibinfo {title}
  {Half-minute-scale atomic coherence and high relative stability in a tweezer
  clock}},\ }\href {https://doi.org/10.1038/s41586-020-3009-y} {\bibfield
  {journal} {\bibinfo  {journal} {Nature}\ }\textbf {\bibinfo {volume} {588}},\
  \bibinfo {pages} {408} (\bibinfo {year} {2020})}\BibitemShut {NoStop}%
\bibitem [{\citenamefont {Ebadi}\ \emph {et~al.}(2022)\citenamefont {Ebadi},
  \citenamefont {Keesling}, \citenamefont {Cain}, \citenamefont {Wang},
  \citenamefont {Levine}, \citenamefont {Bluvstein}, \citenamefont {Semeghini},
  \citenamefont {Omran}, \citenamefont {Liu}, \citenamefont {Samajdar},
  \citenamefont {Luo}, \citenamefont {Nash}, \citenamefont {Gao}, \citenamefont
  {Barak}, \citenamefont {Farhi}, \citenamefont {Sachdev}, \citenamefont
  {Gemelke}, \citenamefont {Zhou}, \citenamefont {Choi}, \citenamefont
  {Pichler}, \citenamefont {Wang}, \citenamefont {Greiner}, \citenamefont
  {Vuleti{\'c}},\ and\ \citenamefont
  {Lukin}}]{ebadiQuantumOptimizationMaximum2022a}%
  \BibitemOpen
  \bibfield  {author} {\bibinfo {author} {\bibfnamefont {S.}~\bibnamefont
  {Ebadi}}, \bibinfo {author} {\bibfnamefont {A.}~\bibnamefont {Keesling}},
  \bibinfo {author} {\bibfnamefont {M.}~\bibnamefont {Cain}}, \bibinfo {author}
  {\bibfnamefont {T.~T.}\ \bibnamefont {Wang}}, \bibinfo {author}
  {\bibfnamefont {H.}~\bibnamefont {Levine}}, \bibinfo {author} {\bibfnamefont
  {D.}~\bibnamefont {Bluvstein}}, \bibinfo {author} {\bibfnamefont
  {G.}~\bibnamefont {Semeghini}}, \bibinfo {author} {\bibfnamefont
  {A.}~\bibnamefont {Omran}}, \bibinfo {author} {\bibfnamefont {J.-G.}\
  \bibnamefont {Liu}}, \bibinfo {author} {\bibfnamefont {R.}~\bibnamefont
  {Samajdar}}, \bibinfo {author} {\bibfnamefont {X.-Z.}\ \bibnamefont {Luo}},
  \bibinfo {author} {\bibfnamefont {B.}~\bibnamefont {Nash}}, \bibinfo {author}
  {\bibfnamefont {X.}~\bibnamefont {Gao}}, \bibinfo {author} {\bibfnamefont
  {B.}~\bibnamefont {Barak}}, \bibinfo {author} {\bibfnamefont
  {E.}~\bibnamefont {Farhi}}, \bibinfo {author} {\bibfnamefont
  {S.}~\bibnamefont {Sachdev}}, \bibinfo {author} {\bibfnamefont
  {N.}~\bibnamefont {Gemelke}}, \bibinfo {author} {\bibfnamefont
  {L.}~\bibnamefont {Zhou}}, \bibinfo {author} {\bibfnamefont {S.}~\bibnamefont
  {Choi}}, \bibinfo {author} {\bibfnamefont {H.}~\bibnamefont {Pichler}},
  \bibinfo {author} {\bibfnamefont {S.-T.}\ \bibnamefont {Wang}}, \bibinfo
  {author} {\bibfnamefont {M.}~\bibnamefont {Greiner}}, \bibinfo {author}
  {\bibfnamefont {V.}~\bibnamefont {Vuleti{\'c}}},\ and\ \bibinfo {author}
  {\bibfnamefont {M.~D.}\ \bibnamefont {Lukin}},\ }\bibfield  {title} {\emph
  {\bibinfo {title} {Quantum optimization of maximum independent set using
  {{Rydberg}} atom arrays}},\ }\href {https://doi.org/10.1126/science.abo6587}
  {\bibfield  {journal} {\bibinfo  {journal} {Science}\ }\textbf {\bibinfo
  {volume} {376}},\ \bibinfo {pages} {1209} (\bibinfo {year}
  {2022})}\BibitemShut {NoStop}%
\bibitem [{\citenamefont {Byun}\ \emph {et~al.}(2022)\citenamefont {Byun},
  \citenamefont {Kim},\ and\ \citenamefont
  {Ahn}}]{byunFindingMaximumIndependent2022}%
  \BibitemOpen
  \bibfield  {author} {\bibinfo {author} {\bibfnamefont {A.}~\bibnamefont
  {Byun}}, \bibinfo {author} {\bibfnamefont {M.}~\bibnamefont {Kim}},\ and\
  \bibinfo {author} {\bibfnamefont {J.}~\bibnamefont {Ahn}},\ }\bibfield
  {title} {\emph {\bibinfo {title} {Finding the {{Maximum Independent Sets}} of
  {{Platonic Graphs Using Rydberg Atoms}}}},\ }\href
  {https://doi.org/10.1103/PRXQuantum.3.030305} {\bibfield  {journal} {\bibinfo
   {journal} {PRX Quantum}\ }\textbf {\bibinfo {volume} {3}},\ \bibinfo {pages}
  {030305} (\bibinfo {year} {2022})}\BibitemShut {NoStop}%
\bibitem [{\citenamefont {Singh}\ \emph {et~al.}(2022)\citenamefont {Singh},
  \citenamefont {Anand}, \citenamefont {Pocklington}, \citenamefont {Kemp},\
  and\ \citenamefont {Bernien}}]{singhDualElementTwoDimensionalAtom2022}%
  \BibitemOpen
  \bibfield  {author} {\bibinfo {author} {\bibfnamefont {K.}~\bibnamefont
  {Singh}}, \bibinfo {author} {\bibfnamefont {S.}~\bibnamefont {Anand}},
  \bibinfo {author} {\bibfnamefont {A.}~\bibnamefont {Pocklington}}, \bibinfo
  {author} {\bibfnamefont {J.~T.}\ \bibnamefont {Kemp}},\ and\ \bibinfo
  {author} {\bibfnamefont {H.}~\bibnamefont {Bernien}},\ }\bibfield  {title}
  {\emph {\bibinfo {title} {Dual-{{Element}}, {{Two-Dimensional Atom Array}}
  with {{Continuous-Mode Operation}}}},\ }\href
  {https://doi.org/10.1103/PhysRevX.12.011040} {\bibfield  {journal} {\bibinfo
  {journal} {Physical Review X}\ }\textbf {\bibinfo {volume} {12}},\ \bibinfo
  {pages} {011040} (\bibinfo {year} {2022})}\BibitemShut {NoStop}%
\bibitem [{\citenamefont {Steinert}\ \emph {et~al.}(2022)\citenamefont
  {Steinert}, \citenamefont {Osterholz}, \citenamefont {Eberhard},
  \citenamefont {Festa}, \citenamefont {Lorenz}, \citenamefont {Chen},
  \citenamefont {Trautmann},\ and\ \citenamefont
  {Gross}}]{steinertSpatiallyProgrammableSpin2022_3}%
  \BibitemOpen
  \bibfield  {author} {\bibinfo {author} {\bibfnamefont {L.-M.}\ \bibnamefont
  {Steinert}}, \bibinfo {author} {\bibfnamefont {P.}~\bibnamefont {Osterholz}},
  \bibinfo {author} {\bibfnamefont {R.}~\bibnamefont {Eberhard}}, \bibinfo
  {author} {\bibfnamefont {L.}~\bibnamefont {Festa}}, \bibinfo {author}
  {\bibfnamefont {N.}~\bibnamefont {Lorenz}}, \bibinfo {author} {\bibfnamefont
  {Z.}~\bibnamefont {Chen}}, \bibinfo {author} {\bibfnamefont {A.}~\bibnamefont
  {Trautmann}},\ and\ \bibinfo {author} {\bibfnamefont {C.}~\bibnamefont
  {Gross}},\ }\bibfield  {title} {\emph {\bibinfo {title} {Spatially
  programmable spin interactions in neutral atom arrays}},\ }\Eprint
  {https://arxiv.org/abs/2206.12385} {arXiv:2206.12385}  (\bibinfo {year}
  {2022})\BibitemShut {NoStop}%
\bibitem [{\citenamefont {Jandura}\ and\ \citenamefont
  {Pupillo}(2022)}]{janduraTimeOptimalTwoThreeQubit2022}%
  \BibitemOpen
  \bibfield  {author} {\bibinfo {author} {\bibfnamefont {S.}~\bibnamefont
  {Jandura}}\ and\ \bibinfo {author} {\bibfnamefont {G.}~\bibnamefont
  {Pupillo}},\ }\bibfield  {title} {\emph {\bibinfo {title} {Time-{{Optimal
  Two-}} and {{Three-Qubit Gates}} for {{Rydberg Atoms}}}},\ }\href
  {https://doi.org/10.22331/q-2022-05-13-712} {\bibfield  {journal} {\bibinfo
  {journal} {Quantum}\ }\textbf {\bibinfo {volume} {6}},\ \bibinfo {pages}
  {712} (\bibinfo {year} {2022})}\BibitemShut {NoStop}%
\bibitem [{\citenamefont {Pagano}\ \emph {et~al.}(2022)\citenamefont {Pagano},
  \citenamefont {Weber}, \citenamefont {Jaschke}, \citenamefont {Pfau},
  \citenamefont {Meinert}, \citenamefont {Montangero},\ and\ \citenamefont
  {B{\"u}chler}}]{paganoErrorbudgetingControlledphaseGate2022_2}%
  \BibitemOpen
  \bibfield  {author} {\bibinfo {author} {\bibfnamefont {A.}~\bibnamefont
  {Pagano}}, \bibinfo {author} {\bibfnamefont {S.}~\bibnamefont {Weber}},
  \bibinfo {author} {\bibfnamefont {D.}~\bibnamefont {Jaschke}}, \bibinfo
  {author} {\bibfnamefont {T.}~\bibnamefont {Pfau}}, \bibinfo {author}
  {\bibfnamefont {F.}~\bibnamefont {Meinert}}, \bibinfo {author} {\bibfnamefont
  {S.}~\bibnamefont {Montangero}},\ and\ \bibinfo {author} {\bibfnamefont
  {H.~P.}\ \bibnamefont {B{\"u}chler}},\ }\bibfield  {title} {\emph {\bibinfo
  {title} {Error-budgeting for a controlled-phase gate with strontium-88
  {{Rydberg}} atoms}},\ }\href
  {https://doi.org/10.1103/PhysRevResearch.4.033019} {\bibfield  {journal}
  {\bibinfo  {journal} {Physical Review Research}\ }\textbf {\bibinfo {volume}
  {4}},\ \bibinfo {pages} {033019} (\bibinfo {year} {2022})}\BibitemShut
  {NoStop}%
\bibitem [{\citenamefont {Saffman}\ \emph {et~al.}(2010)\citenamefont
  {Saffman}, \citenamefont {Walker},\ and\ \citenamefont
  {M\o{}lmer}}]{saffmannQuantumInformationRydberg2010_2}%
  \BibitemOpen
  \bibfield  {author} {\bibinfo {author} {\bibfnamefont {M.}~\bibnamefont
  {Saffman}}, \bibinfo {author} {\bibfnamefont {T.~G.}\ \bibnamefont
  {Walker}},\ and\ \bibinfo {author} {\bibfnamefont {K.}~\bibnamefont
  {M\o{}lmer}},\ }\bibfield  {title} {\emph {\bibinfo {title} {Quantum
  information with {{Rydberg}} atoms}},\ }\href
  {https://doi.org/10.1103/RevModPhys.82.2313} {\bibfield  {journal} {\bibinfo
  {journal} {Reviews of Modern Physics}\ }\textbf {\bibinfo {volume} {82}},\
  \bibinfo {pages} {2313} (\bibinfo {year} {2010})}\BibitemShut {NoStop}%
\bibitem [{\citenamefont {Vandersypen}\ and\ \citenamefont
  {Chuang}(2005)}]{VandersypenNMRtechniques}%
  \BibitemOpen
  \bibfield  {author} {\bibinfo {author} {\bibfnamefont {L.~M.~K.}\
  \bibnamefont {Vandersypen}}\ and\ \bibinfo {author} {\bibfnamefont {I.~L.}\
  \bibnamefont {Chuang}},\ }\bibfield  {title} {\emph {\bibinfo {title}
  {{{NMR}} techniques for quantum control and computation}},\ }\href
  {https://doi.org/https://doi.org/10.1103/RevModPhys.76.1037} {\bibfield
  {journal} {\bibinfo  {journal} {Reviews of Modern Physics}\ }\textbf
  {\bibinfo {volume} {76}},\ \bibinfo {pages} {1037} (\bibinfo {year}
  {2005})}\BibitemShut {NoStop}%
\bibitem [{\citenamefont {Gullion}\ \emph {et~al.}(1990)\citenamefont
  {Gullion}, \citenamefont {Baker},\ and\ \citenamefont
  {Conradi}}]{GullionCarrPurcell}%
  \BibitemOpen
  \bibfield  {author} {\bibinfo {author} {\bibfnamefont {T.}~\bibnamefont
  {Gullion}}, \bibinfo {author} {\bibfnamefont {D.~B.}\ \bibnamefont {Baker}},\
  and\ \bibinfo {author} {\bibfnamefont {M.~S.}\ \bibnamefont {Conradi}},\
  }\bibfield  {title} {\emph {\bibinfo {title} {New, compensated
  {{Carr-Purcell}} sequences}},\ }\href
  {https://doi.org/https://doi.org/10.1016/0022-2364(90)90331-3} {\bibfield
  {journal} {\bibinfo  {journal} {Journal of Magnetic Resonance}\ }\textbf
  {\bibinfo {volume} {89}},\ \bibinfo {pages} {479} (\bibinfo {year}
  {1990})}\BibitemShut {NoStop}%
\bibitem [{\citenamefont {Pedersen}\ \emph {et~al.}(2007)\citenamefont
  {Pedersen}, \citenamefont {M{\o}ller},\ and\ \citenamefont
  {M{\o}lmer}}]{Pedersen_2007}%
  \BibitemOpen
  \bibfield  {author} {\bibinfo {author} {\bibfnamefont {L.~H.}\ \bibnamefont
  {Pedersen}}, \bibinfo {author} {\bibfnamefont {N.~M.}\ \bibnamefont
  {M{\o}ller}},\ and\ \bibinfo {author} {\bibfnamefont {K.}~\bibnamefont
  {M{\o}lmer}},\ }\bibfield  {title} {\emph {\bibinfo {title} {Fidelity of
  quantum operations}},\ }\href
  {https://doi.org/10.1016/j.physleta.2007.02.069} {\bibfield  {journal}
  {\bibinfo  {journal} {Physics Letters A}\ }\textbf {\bibinfo {volume}
  {367}},\ \bibinfo {pages} {47} (\bibinfo {year} {2007})}\BibitemShut
  {NoStop}%
\bibitem [{\citenamefont {Shapira}\ \emph {et~al.}(2018)\citenamefont
  {Shapira}, \citenamefont {Shaniv}, \citenamefont {Manovitz}, \citenamefont
  {Akerman},\ and\ \citenamefont {Ozeri}}]{shapiraRobustIon}%
  \BibitemOpen
  \bibfield  {author} {\bibinfo {author} {\bibfnamefont {Y.}~\bibnamefont
  {Shapira}}, \bibinfo {author} {\bibfnamefont {R.}~\bibnamefont {Shaniv}},
  \bibinfo {author} {\bibfnamefont {T.}~\bibnamefont {Manovitz}}, \bibinfo
  {author} {\bibfnamefont {N.}~\bibnamefont {Akerman}},\ and\ \bibinfo {author}
  {\bibfnamefont {R.}~\bibnamefont {Ozeri}},\ }\bibfield  {title} {\emph
  {\bibinfo {title} {Robust {{Entanglement Gates}} for {{Trapped-Ion
  Qubits}}}},\ }\href {https://doi.org/10.1103/PhysRevLett.121.180502}
  {\bibfield  {journal} {\bibinfo  {journal} {Physical Review Letters}\
  }\textbf {\bibinfo {volume} {121}},\ \bibinfo {pages} {180502} (\bibinfo
  {year} {2018})}\BibitemShut {NoStop}%
\bibitem [{\citenamefont {Wu}\ \emph {et~al.}(2022)\citenamefont {Wu},
  \citenamefont {Kolkowitz}, \citenamefont {Puri},\ and\ \citenamefont
  {Thompson}}]{Wu2022NatComm}%
  \BibitemOpen
  \bibfield  {author} {\bibinfo {author} {\bibfnamefont {Y.}~\bibnamefont
  {Wu}}, \bibinfo {author} {\bibfnamefont {S.}~\bibnamefont {Kolkowitz}},
  \bibinfo {author} {\bibfnamefont {S.}~\bibnamefont {Puri}},\ and\ \bibinfo
  {author} {\bibfnamefont {J.~D.}\ \bibnamefont {Thompson}},\ }\bibfield
  {title} {\emph {\bibinfo {title} {Erasure conversion for fault-tolerant
  quantum computing in alkaline earth {{Rydberg}} atom arrays}},\ }\href
  {https://doi.org/10.1038/s41467-022-32094-6} {\bibfield  {journal} {\bibinfo
  {journal} {Nature Communications}\ }\textbf {\bibinfo {volume} {13}},\
  \bibinfo {pages} {4657} (\bibinfo {year} {2022})}\BibitemShut {NoStop}%
\bibitem [{\citenamefont {Graham}\ \emph {et~al.}(2019)\citenamefont {Graham},
  \citenamefont {Kwon}, \citenamefont {Grinkemeyer}, \citenamefont {Marra},
  \citenamefont {Jiang}, \citenamefont {Lichtman}, \citenamefont {Sun},
  \citenamefont {Ebert},\ and\ \citenamefont
  {Saffman}}]{grahamRydbergMediatedEntanglementTwoDimensional2019}%
  \BibitemOpen
  \bibfield  {author} {\bibinfo {author} {\bibfnamefont {T.~M.}\ \bibnamefont
  {Graham}}, \bibinfo {author} {\bibfnamefont {M.}~\bibnamefont {Kwon}},
  \bibinfo {author} {\bibfnamefont {B.}~\bibnamefont {Grinkemeyer}}, \bibinfo
  {author} {\bibfnamefont {Z.}~\bibnamefont {Marra}}, \bibinfo {author}
  {\bibfnamefont {X.}~\bibnamefont {Jiang}}, \bibinfo {author} {\bibfnamefont
  {M.~T.}\ \bibnamefont {Lichtman}}, \bibinfo {author} {\bibfnamefont
  {Y.}~\bibnamefont {Sun}}, \bibinfo {author} {\bibfnamefont {M.}~\bibnamefont
  {Ebert}},\ and\ \bibinfo {author} {\bibfnamefont {M.}~\bibnamefont
  {Saffman}},\ }\bibfield  {title} {\emph {\bibinfo {title} {Rydberg-{{Mediated
  Entanglement}} in a {{Two-Dimensional Neutral Atom Qubit Array}}}},\ }\href
  {https://doi.org/10.1103/PhysRevLett.123.230501} {\bibfield  {journal}
  {\bibinfo  {journal} {Physical Review Letters}\ }\textbf {\bibinfo {volume}
  {123}},\ \bibinfo {pages} {230501} (\bibinfo {year} {2019})}\BibitemShut
  {NoStop}%
\bibitem [{Note1()}]{Note1}%
  \BibitemOpen
  \bibinfo {note} {This works as long as the uncontrolled phase shifts acquired
  in the Doppler inversion are small, as in typical experimental setups
  (Appendix~\ref {App:DopplerEcho}). If they are large, then the Doppler
  inversion should only be applied after the application of leakage robust
  sequences.}\BibitemShut {Stop}%
\bibitem [{\citenamefont {Jandura}\ \emph {et~al.}(2022)\citenamefont
  {Jandura}, \citenamefont {Thompson},\ and\ \citenamefont
  {Pupillo}}]{janduraOptimizingGates}%
  \BibitemOpen
  \bibfield  {author} {\bibinfo {author} {\bibfnamefont {S.}~\bibnamefont
  {Jandura}}, \bibinfo {author} {\bibfnamefont {J.~D.}\ \bibnamefont
  {Thompson}},\ and\ \bibinfo {author} {\bibfnamefont {G.}~\bibnamefont
  {Pupillo}},\ }\bibfield  {title} {\emph {\bibinfo {title} {Optimizing
  {{Rydberg Gates}} for {{Logical Qubit Performance}}}},\ }\Eprint
  {https://arxiv.org/abs/2210.06879} {arXiv:2210.06879}  (\bibinfo {year}
  {2022})\BibitemShut {NoStop}%
\bibitem [{\citenamefont {Kaufman}\ \emph {et~al.}(2012)\citenamefont
  {Kaufman}, \citenamefont {Lester},\ and\ \citenamefont
  {Regal}}]{kaufmanCoolingSingleAtom}%
  \BibitemOpen
  \bibfield  {author} {\bibinfo {author} {\bibfnamefont {A.~M.}\ \bibnamefont
  {Kaufman}}, \bibinfo {author} {\bibfnamefont {B.~J.}\ \bibnamefont
  {Lester}},\ and\ \bibinfo {author} {\bibfnamefont {C.~A.}\ \bibnamefont
  {Regal}},\ }\bibfield  {title} {\emph {\bibinfo {title} {Cooling a {{Single
  Atom}} in an {{Optical Tweezer}} to {{Its Quantum Ground State}}}},\ }\href
  {https://doi.org/10.1103/PhysRevX.2.041014} {\bibfield  {journal} {\bibinfo
  {journal} {Physical Review X}\ }\textbf {\bibinfo {volume} {2}},\ \bibinfo
  {pages} {041014} (\bibinfo {year} {2012})}\BibitemShut {NoStop}%
\bibitem [{\citenamefont {Zhang}\ \emph {et~al.}(2011)\citenamefont {Zhang},
  \citenamefont {Robicheaux},\ and\ \citenamefont
  {Saffman}}]{zhangMagictrapping}%
  \BibitemOpen
  \bibfield  {author} {\bibinfo {author} {\bibfnamefont {S.}~\bibnamefont
  {Zhang}}, \bibinfo {author} {\bibfnamefont {F.}~\bibnamefont {Robicheaux}},\
  and\ \bibinfo {author} {\bibfnamefont {M.}~\bibnamefont {Saffman}},\
  }\bibfield  {title} {\emph {\bibinfo {title} {Magic-wavelength optical traps
  for {{Rydberg}} atoms}},\ }\href {https://doi.org/10.1103/PhysRevA.84.043408}
  {\bibfield  {journal} {\bibinfo  {journal} {Physical Review A}\ }\textbf
  {\bibinfo {volume} {84}},\ \bibinfo {pages} {043408} (\bibinfo {year}
  {2011})}\BibitemShut {NoStop}%
\end{thebibliography}%

\end{document}